\newif\ifarxiv
\arxivtrue
\documentclass[conference,compsoc]{IEEEtran}
\PassOptionsToPackage{hyphens}{url}

\newcommand{\tool}{\textsc{HqcLethe}\xspace}
\newcommand{\alg}[1]{\textsc{#1}}
\newcommand{\pkval}[1]{\textcolor{blue}{#1}}
\newcommand{\encval}[1]{\textcolor{red}{#1}}
\newcommand{\Fq}{\mathbb{F}}
\newcommand{\artifacturl}{%
  \ifarxiv
    \url{https://github.com/PLSec-VU/hqctiny}%
  \else
    \url{https://anonymous.4open.science/r/hqctiny-1C73/}%
  \fi}

\usepackage{url}
\usepackage{tikz}
\usetikzlibrary{arrows.meta,decorations.pathreplacing,calc,positioning}
\usepackage{pgfplots}
\pgfplotsset{compat=1.16}
\usetikzlibrary{patterns}
\definecolor{hqclethecolor}{HTML}{2A78D6}
\definecolor{statefulcolor}{HTML}{EB6834}
\definecolor{mctinycolor}{HTML}{1BAF7A}
\definecolor{inkmuted}{HTML}{52514E}
\pgfplotsset{
  evalplot/.style={
    font=\scriptsize, tick label style={font=\scriptsize}, label style={font=\scriptsize},
    title style={font=\scriptsize, yshift=-4pt}, legend style={font=\tiny, draw=none, fill=white, fill opacity=0.85, text opacity=1},
    axis line style={inkmuted}, tick style={inkmuted}, major grid style={gray!20}, ymajorgrids,
    legend cell align=left, every axis plot/.append style={line width=0.8pt},
    error bars/error bar style={line width=0.5pt, solid}, error bars/error mark options={rotate=90, mark size=1.5pt},
  },
  hqclethe/.style={hqclethecolor, solid, mark=*, mark size=1.6pt, mark options={fill=hqclethecolor}},
  stateful/.style={statefulcolor, densely dashed, mark=square*, mark size=1.5pt, mark options={solid, fill=statefulcolor}},
  mctiny/.style={mctinycolor, densely dotted, mark=triangle*, mark size=2pt, mark options={solid, fill=mctinycolor}},
  hqclethebar/.style={draw=none, fill=hqclethecolor, area legend},
  statefulbar/.style={draw=none, fill=statefulcolor, area legend, postaction={pattern=north east lines, pattern color=white}},
  statefulone/.style={statefulcolor, densely dotted, thick},
  statefulthree/.style={statefulcolor, densely dashed, thick},
  statefulfive/.style={statefulcolor, solid, thick},
  measured/.style={only marks, mark=*, mark size=1.6pt, black},
  hqcletheband/.style={draw=none, fill=hqclethecolor, fill opacity=0.35},
  pi ram/.style={inkmuted, densely dashdotted, thin},
}
\tikzset{
  statelabel/.style={anchor=west, font=\tiny, text=inkmuted},
  ratelabel/.style={font=\tiny, text=inkmuted},
  hqclethelabel/.style={anchor=north west, font=\tiny, text=hqclethecolor!70!black},
  ramlabel/.style={anchor=south west, font=\tiny, text=inkmuted},
  floodsplit/.style={inkmuted, thin},
}

\usepackage{amsmath,amssymb}
\usepackage{xcolor}
\usepackage{array}
\usepackage{booktabs}
\usepackage{float}
\usepackage{algorithm}
\usepackage{algpseudocode}
\usepackage{xspace}
\usepackage{textcomp}
\ifarxiv
  \usepackage[hidelinks]{hyperref}
\fi
\usepackage{cleveref}

\begin{document}

\title{\tool: Streaming HQC Encapsulation\\
  for Tiny Stateless Servers}

\ifarxiv
  \author{
    \IEEEauthorblockN{Wind~Wong, Klaus~v.~Gleissenthall}
    \IEEEauthorblockA{\{t.f.w.wong, kfl900\}@vu.nl\\
      Vrije Universiteit Amsterdam}
  }
\else
  \author{\IEEEauthorblockN{Anonymous Submission}}
\fi

\maketitle

\begin{abstract}
Stateless post-quantum authenticated key exchanges such as post-quantum
WireGuard hedge against unexpected cryptanalytic advances by combining a lattice
KEM with code-based Classic McEliece. However, recent attacks have led Germany's
BSI to advise against McEliece, which leaves these protocols with lattices
alone. HQC, the code-based KEM that NIST selected, could restore diversity, but
its keys do not fit in a single packet, so a server would have to buffer
multi-packet handshakes. This would leave an attack surface for memory
exhaustion, which is what WireGuard's stateless handshake was designed to avoid
in the first place.
McTiny demonstrated that stateless key encapsulation for Classic McEliece is
possible by streaming key blocks one-at-a-time.
However, HQC's encapsulation does not split as easily into individual packets,
as cyclic polynomial multiplication, unlike matrix-vector multiplication, is
inherently non-local.

We present \tool, a stateless key exchange that streams an HQC encapsulation in
packet-sized pieces. It uses the Chinese remainder theorem to split the
polynomial multiplication into independent remainders, one per packet, and masks
them so that they reveal no more than an ordinary HQC ciphertext, which we prove
in Rocq with SSProve; we also verify session secrecy, server authentication and
forward secrecy in ProVerif. Client puzzles bound CPU exhaustion, and \tool
removes memory exhaustion: on a Raspberry Pi~3B+, each worker runs in a fixed
1.2--1.6\,MiB. At category~5 it sends $27\times$ fewer bytes and uses
$9.0\times$ less server CPU than McTiny, the earlier stateless code-based
design.
\end{abstract}

\section{Introduction}
\label{sec:intro}

Authenticated key exchange (AKE) lets two parties establish a shared secret while
verifying each other's identities. One common design for security against
quantum adversaries uses two post-quantum key-encapsulation mechanisms (KEMs) in
each session: a long-term, static KEM key pair for authentication, and a fresh,
ephemeral KEM key pair used only for that session. The ephemeral key pair
provides forward secrecy: because its secret key is erased after the exchange, a
later compromise of the long-term key does not reveal recorded past sessions.

A server that wants to remain
cheap to operate---say, a VPN endpoint on a home router, or a gateway on an IoT
device---has very little memory to spare, and every byte it allocates on behalf
of an unauthenticated client is a lever for an attacker. If completing a
handshake requires the server to buffer per-client state, an adversary can open
thousands of half-finished handshakes at once, forcing the server to allocate
memory for each which exhausts the device easily. This type of memory-flooding
denial-of-service attack, is exactly what low-footprint network software is
built to avoid. Indeed, WireGuard~\cite{wireguard} owes much of its adoption,
especially on constrained devices, to a stateless UDP design in which the server
holds no per-client state until a handshake is cryptographically verified. We
call a server with this property a \emph{tiny network server}: it keeps no
per-client state, and its memory is fixed, independent of the number of clients
and of the attacker's effort.

Previous efforts to make WireGuard post-quantum
secure~\cite{pq-wireguard,pqwireguard_revisited,rosenpass} obtain their
ephemeral security from lattices, and reserve code-based cryptography only for
long-lived static keys, with Classic McEliece~\cite{classicmceliece} being the
code-based KEM of choice. This hybrid design stems from a general consensus that
post-quantum cryptanalysis is too new to rely entirely on one mathematical
family alone. Classic McEliece was considered to be uniquely suited for long
term keys, because it has a long history dating to McEliece's 1978
proposal~\cite{mceliece}, and was the oldest and most conservatively studied
post-quantum family~\cite{mojaveri2026mceliece}. While its keys are large: the
Classic McEliece parameters recommended for long-term security use public keys
of roughly 1\,MB, even the largest parameters have ciphertexts of only 208
bytes, allowing it to fit comfortably within UDP packets, retaining
statelessness.
However, attacks published in 2026 have called Classic McEliece's security into
doubt: building on the distinguisher of Ghoshal et al.~\cite{GIJS2026}, Weis
claims key recovery at $2^{99}$ to $2^{102}$ operations for the recommended
parameter sets~\cite{Weis2026,saarinen2026cost,saarinen2026forum}. On 29 September 2026,
Germany's BSI advised that ``Classic McEliece should currently no longer be used
in new developments or when planning new cryptographic applications''~\cite{BSI2026}.
A code-based replacement for Classic McEliece is needed.

HQC is the natural candidate. In March 2025, NIST announced that ``the only
key-establishment algorithm that will be standardized is
HQC''~\cite{nist_ir8545,nist-hqc-selection}, selected from candidates meant to
``reduce the risk that a single cryptanalytic breakthrough will leave no viable
standard for key establishment''. The BSI note finds HQC unaffected by the new
attacks and calls it ``an important alternative to the lattice-based
post-quantum schemes''~\cite{BSI2026}. Its public keys are far smaller than
Classic McEliece's---about 2.2\,KB, 4.5\,KB, and 7.2\,KB at its three security
levels---small enough to make a code-based ephemeral KEM practical, yet still
several packets wide, and the ciphertexts are twice as big. A 2--7\,KB key and
4--14\,KB ciphertext do not fit in the 1280-byte payload that IPv6 guarantees
will not fragment~\cite{rfc8200}, and it is precisely the size at which an
implementor is tempted by the simplest and seemingly obvious fix: split the
content across a few datagrams, store them in a bounded buffer until they all
arrive, and proceed. However, such bounds are deceptively hard to get right: in
HTTP/2, attackers exhausted servers with gigabytes of memory through unbounded
header fragments~\cite{cert-continuation}. Similarly, streams were counted as
closed while their work continued~\cite{madeyoureset}. A better solution is to
remove this attack surface altogether.

McTiny~\cite{mctiny} showed the obstacle is not insurmountable.  McTiny runs a
Classic McEliece key exchange on a tiny network server, by having the client
split its key into row/column blocks small enough for a packet and having the
server compute and forget one partial matrix--vector product per block, never
assembling the whole key. 
%
%
Unfortunately, HQC's encapsulation is cyclic polynomial arithmetic, which,
unlike matrix--vector multiplication, is inherently non-local.

In this paper, we introduce \tool, the first tiny-network-server realization of
HQC key encapsulation. Our key idea is to use the Chinese remainder theorem
(CRT) to break polynomial multiplication into independent computations on
residues: the client sends the public key polynomial modulo packet-sized,
pairwise coprime polynomials, and the server computes and returns the
corresponding ciphertext residue without retaining per-client state. The client
then reconstructs the ciphertext from the returned residues. Since this method
produces unreduced polynomial products whereas plain HQC ciphertexts are reduced
against HQC's ring modulus, \tool masks the residues to prevent them from
leaking information about the server's secret randomness.
We build an AKE protocol around it in which HQC provides the ephemeral key and
ML-KEM-768 the authentication, keeping the code and lattice diversity of prior
work without Classic McEliece.

The result is a post-quantum AKE with forward secrecy by NIST chosen KEMs,
two-round-trip completion, and no per-client server state: self-encrypted
cookies carry session state across queries (\S\ref{sec:protocol}). Of the three
attack surfaces for denial of service, \tool removes memory exhaustion entirely.
CPU exhaustion, which any server that authenticates its first message suffers,
is bounded by standard proof-of-work client puzzles that tie an attacker's
request rate to its CPU (\S\ref{sec:eval-dos}, Appendix~\ref{app:puzzle-cost}),
and saturating the network link is left to filtering upstream, as for any
protocol. Beyond the protocol, we contribute masked CRT-based residue streaming,
with a Rocq/SSProve proof that the stream achieves zero-leakage simulation
against an ordinary HQC ciphertext, and a ProVerif proof of session secrecy,
server authentication, and the protocol's forward-secrecy boundary under active
key compromise (\S\ref{sec:mask}, \S\ref{sec:formal}, \S\ref{sec:proverif}). Our
Rust implementation, with AVX2 and NEON acceleration, sends $27\times$ fewer
bytes and $44\times$ fewer packets than McTiny at category~5, while using
$9.5\times$ less server CPU on one AVX2 core ($9.0\times$ less on a Raspberry
Pi~3B+); most of the savings come from HQC's smaller keys
(\S\ref{sec:evaluation}). Compared with a stateful buffering baseline, \tool
uses $2.0\times$ the bytes and $4.1\times$ the CPU, but each worker's memory
remains fixed at 1.2--1.6\,MiB on a Raspberry Pi, independent of attacker
activity. A stateful server's memory instead grows with the attacker: even with
the same client puzzle, each attacker core can hold up to 8.5\,MB of its table
at category~5 (\S\ref{sec:eval-dos}). \tool trades moderate CPU overhead for
memory that no attacker can exhaust.
\section{Technical Overview}
\label{sec:overview}

We illustrate the core technique using a minimal polynomial example;
\S\ref{sec:keys} presents the full-scale construction.

\begin{figure}[tp]
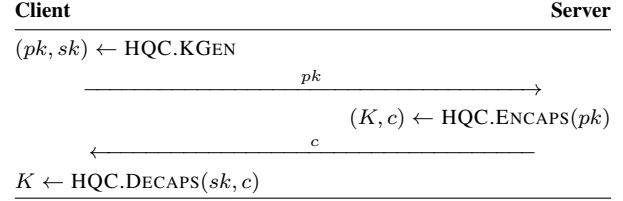

\centering
\footnotesize
\begin{tabular}{@{}p{0.46\linewidth}@{\extracolsep{\fill}}>{\raggedleft\arraybackslash}p{0.46\linewidth}@{}}
\textbf{Client} & \textbf{Server} \\[2pt]
\hline\noalign{\smallskip}
$(pk,sk)\gets \alg{HQC.KGen}$ & \\[2pt]
\multicolumn{2}{c}{$\xrightarrow{\hspace*{2.8cm} pk \hspace*{2.8cm}}$} \\[2pt]
 & $(K,c)\gets \alg{HQC.Encaps}(pk)$ \\[2pt]
\multicolumn{2}{c}{$\xleftarrow{\hspace*{2.8cm} c \hspace*{2.8cm}}$} \\[2pt]
$K\gets \alg{HQC.Decaps}(sk,c)$ & \\[1pt]
\hline
\end{tabular}
\caption{A strawman of a key exchange protocol.}
\label{fig:hqc-strawman}
\end{figure}

A client wants to run key encapsulation with a server, against a
fresh public key the client just generated, as shown in
Figure~\ref{fig:hqc-strawman}.
The client creates a new secret/public key pair and sends the public key to the
server. The server runs encapsulation, which creates an ephemeral session key
$K$, and an encapsulation~$c$. The server sends the encapsulation---an encrypted
version of key~$K$---to the client who decrypts it using secret key~$sk$ to
receive the same shared key~$K$. The client then destroys~$sk$; from now on, no
recorded traffic using~$K$ can be decrypted by an adversary
(\S\ref{sec:erasure}). We want to run this on a tiny network server (\S\ref{sec:intro}). The obstacle
is the size of the public key: an HQC public key is 2.2--7.2\,KB
(\S\ref{sec:codecrypto}), several packets' worth, and buffering its fragments
would expose the server to memory exhaustion. As
such, public key~$pk$ may only ever visit the server in packet-sized pieces, and
the server must treat every piece as if it were the first: do a packet-sized
piece of the encryption, send back a packet-sized piece of the answer, forget
everything. The question is whether we can soundly partition HQC encryption into
suitable packet-sized chunks.

\emph{Multiplication is Non-Local.} Stripping away detail, HQC encryption boils down
to multiplying and adding polynomials whose coefficients are single bits, added
without carries ($1+1=0$; bit-wise XOR). One factor, let's call it $p$, comes
from the client's public key; two others, $q,\,r$, are secret randomness held by
the server; both have thousands of coefficients, and the ciphertext is $pq + r$.
For McTiny~\cite{mctiny} the
corresponding operation was friendlier: Classic McEliece encryption multiplies a
\emph{matrix} (the key) by a vector, and each block of the key contributes to a
bounded part of the output, so the server can process one block per
packet and forget. Cyclic multiplication of polynomials is different.
Consider $p(X)\cdot q(X)$ for $p(X)=X^3$ and $q(X)=1+X+X^2$ in
the 4-bit ring $F_2[X]/(X^4-1)$. Schoolbook multiplication asks us to
multiply every term of $q(X)$ by every term of $p(X)$: $X^3\cdot 1 + X^3\cdot X
+ X^3\cdot X^2 = X^3+X^4+X^5$. Worse, cyclic multiplication wraps coefficients above degree
$3$ back around: since $X^4=1$, this reduces to $1+X+X^3$. A single term of the
public key has reached coefficients across the whole output, and in general any term
of $p(X)$ may affect any term of the result.
While one could represent cyclic multiplication by $p$ as multiplication by a
matrix whose rows are cyclic shifts of $p$, and use McTiny-style block processing,
the resulting matrix would be prohibitively large.

\emph{Using the Chinese Remainder Theorem.}
The Chinese remainder theorem (CRT) represents a number or polynomial by its
residues (remainders) modulo coprime moduli, then reconstructs it from those
residues. Consider an example that follows HQC's encryption but with numbers
instead of polynomials: the client wishes to send $p$ to the server, and the
server wishes to return $n = pq+r$ without revealing $q,r$. The chosen public
parameter bounds $0 \le n < 42$, but packets can only fit in numbers of at most
7. The CRT lets us solve this problem by having the client send $x = p \bmod
m_i$ and the server responds with $x\cdot q+r \bmod m_i$, with the $m_i$'s being
coprime. Suppose $p,q,r = 9,4,3$ respectively, and $m_{1,2} = \{6,7\}$. The
client sends $9 \bmod 6 = 3$ and $9 \bmod 7 = 2$. The server then computes the
results locally per residue: it responds with $3\cdot4 + 3 \bmod 6 = 3$ and
$2\cdot4+3 \bmod 7 = 4$ respectively. The packets $3,2,3,4$ all fit within the
packet limit, and the server can forget the client's packet immediately after
computing the response. After collecting the responses, the client obtains the
relation $n = 3 + 6a$, and $n = 4 + 7b$. By the CRT, there is only one possible
solution for $a, b$ where $0 \le n < 42$, giving us $n = 39$, which matches the
direct result~$9\cdot4+3$.

\emph{What's missing?} Two complications remain. First, the server must use the
\emph{same} randomness $q,r$ at every multiplication step for each session, and
$q,r$ must be unique per-session, yet remember nothing between packets; \tool
resolves this with a cookie in the tradition of McTiny~\cite{mctiny}: at
handshake time the server encrypts the session's secrets \emph{to itself} under
a key only it holds, and every later request carries the cookie back so the
server can reopen it, rederive $q,r$, and forget again (\S\ref{sec:cookies}).
Second, the point stream \emph{overshares}: the above method lets the server
compute and transmit $pq+r$ statelessly, but a plain HQC ciphertext is not
$pq+r$ itself but rather the residue $pq+r \bmod (X^n - 1)$, and exposing the
unreduced product-sum reveals secrets to attackers.  \tool adds
random masks on the server side to cancel this surplus (\S\ref{sec:mask}).
\section{Code-Based Cryptography and HQC}
\label{sec:codecrypto}

The overview treated HQC encryption abstractly as adding and multiplying
polynomials over $F_2$. We now discuss it in more detail.

\emph{Code-based cryptography} bases its security on the difficulty of decoding
a random-looking linear code, a problem structurally
unrelated to the lattice problems behind ML-KEM~\cite{alagic2024module}, so advances against lattices
would not touch it.

HQC (Hamming Quasi-Cyclic)~\cite{8288618}, selected by NIST instead~\cite{hqc,nist-hqc-selection},
achieves the same security levels
while shrinking keys by three orders of magnitude. It does so by working in the
\emph{quasi-cyclic ring}: $R = F_2[X]/(X^n-1)$,
whose elements are binary polynomials of degree ${<}n$, or equivalently $n$-bit
vectors. Multiplication in $R$ is cyclic convolution, so $X^{n}=1$. An HQC
public key is a pair $(h,s)\in R^2$; $h$ is stored compactly as a short seed and
expanded on demand, so the transmitted key is essentially the single ring
element $s$ plus the seed. The matching secret key is a pair of sparse ring
elements. Three parameter sets tuned to NIST's security categories 1,
3, and 5:

\begin{center}
\begin{tabular}{lrrr}
\hline
 & HQC-1 & HQC-3 & HQC-5 \\
\hline
Codeword size $n_1 n_2$ & 17\,664 & 35\,840 & 57\,600 \\
Ring degree $n$      & 17\,669 & 35\,851 & 57\,637 \\
Public key            & 2\,241\,B & 4\,514\,B & 7\,237\,B \\
Ciphertext            & 4\,433\,B & 8\,978\,B & 14\,421\,B \\
\hline
\end{tabular}
\end{center}

\noindent
These keys are three orders of magnitude smaller than Classic McEliece's, but
still two orders of magnitude larger than a network packet. The rest of this
paper is about closing that gap without letting a tiny network server buffer a
client's key or ciphertext across packets.

\subsection{HQC Encapsulation}
\label{sec:hqc-encaps}

HQC is a \emph{key encapsulation mechanism} (KEM). Rather than encrypting a
chosen message, one party runs \alg{HQC.Encaps} against another's public key to
produce, in one shot, a fresh shared secret $K$ together with a ciphertext $c$
that transports it. The holder of the secret key recovers the same $K$ by
running \alg{HQC.Decaps} on $c$. \tool's server implements \emph{only} the
encapsulation side, against a client-supplied public key, and streams the
resulting ciphertext back as it is produced; the client runs the trusted
decapsulation routine provided by the reference implementation, at the end.
\alg{HQC.Encaps} is therefore central to the rest of the paper, since
\S\ref{sec:protocol} re-expresses Algorithm~\ref{alg:hqc-encaps} as a sequence
of packet-sized steps.

Encapsulation is built in two layers. Underneath sits a public-key encryption
scheme that is secure only against a passive eavesdropper. On top of it, HQC
applies a variant~\cite{fo_modular} of the standard Fujisaki--Okamoto
transform~\cite{fujisaki_okamoto}, which derives all encryption randomness
deterministically from the raw plaintext $m$ and a public salt. After decapsulating
a ciphertext $c$ and obtaining $m'$, encapsulation is re-run with $m'$ and
the salt to reproduce $c'$, which $c$ is checked against to confirm it is a
valid ciphertext, upgrading the scheme to withstand active attackers who
can submit malformed ciphertexts that might compromise $sk$.
The shared secret $K$ is then derived from $m'$, which supplies its secrecy,
together with further values that bind it to the ciphertext and the public key.
We describe both layers together, since \tool reimplements the whole of
Algorithm~\ref{alg:hqc-encaps}.

\begin{algorithm}
\caption{\alg{HQC.Encaps}($\pkval{pk}$)}
\label{alg:hqc-encaps}
\begin{algorithmic}[1]
\Require Public key $\pkval{pk}=(\pkval{h},\pkval{s})$.
\Ensure Shared secret $K$, ciphertext $(u, v, \encval{\mathit{salt}})$.
\State $\encval{m} \xleftarrow{\$} \{0,1\}^{k}$
  \Comment{random seed message}
\State $\encval{\mathit{salt}} \xleftarrow{\$} \{0,1\}^{128}$
\State $(K,\encval{\theta}) \gets G\big(H(\pkval{pk}) \,\|\, \encval{m} \,\|\, \encval{\mathit{salt}}\big)$
  \Comment{$H$ hash, $G$ PRG}
\State $(\encval{r_1},\encval{r_2},\encval{e}) \gets \textsc{SampleFixedWeight}(\encval{\theta})$
\State $\encval{\vec{m}} \gets \textsc{Encode}(\encval{m})$
\State $u \gets \encval{r_1} + \pkval{h}\cdot \encval{r_2}$
\State $v \gets \encval{\vec{m}} + \textsc{Truncate}(\pkval{s}\cdot \encval{r_2} + \encval{e})$
\State \Return $(K, u, v, \encval{\mathit{salt}})$
\end{algorithmic}
\end{algorithm}

\noindent
Writing \pkval{blue} for values that come from the client's public key and
\encval{red} for values the encapsulator freshly samples: $\encval{m}$ is a
fresh random bit string, not a payload the caller chooses --- it seeds
everything else and is recovered by the decapsulator, which re-derives $K$ from
it --- and $\encval{\mathit{salt}}$ is a fresh public random string appended to
the ciphertext in the clear. $H$ is a hash and $G$ a pseudorandom generator
that stretches its input into the shared secret $K$ and a randomness seed
$\encval{\theta}$, from which the sparse fixed-weight noise vectors
$\encval{r_1},\encval{r_2},\encval{e}$ are derived deterministically;
crucially, the public key enters only through its hash $H(\pkval{pk})$. The two
ciphertext ring elements are $u$, in which the public $\pkval{h}$ masks the
noise, and $v$, in which the message is first encoded into a codeword
$\encval{\vec{m}}$ by $\textsc{Encode}$ and then hidden under the noisy term
$\pkval{s}\cdot \encval{r_2}+\encval{e}$, with $\textsc{Truncate}$ discarding
$v$'s coefficients past position $n_1 n_2$. Decapsulation removes the noise and
decodes $\encval{\vec{m}}$ to recover $\encval{m}$.

Three properties of Algorithm~\ref{alg:hqc-encaps} shape the protocol. First,
the public key enters only through its hash: line~3 uses only
$H(\pkval{pk})$, so the server can derive $K$ and $\encval{\theta}$ from that
hash and needs $\pkval{h}$ and $\pkval{s}$ only to form $u$ and $v$
(lines~6--7); the one-packet handshake therefore carries $H(\pkval{pk})$, and
the client streams the key afterwards (\S\ref{sec:protocol}). Second, every
packet-sized piece of $u$ and $v$ needs $\encval{r_1},\encval{r_2},\encval{e}$
and $\encval{\vec{m}}$ again (lines~4--5); a stateless server can neither
resample them, which would break correctness, nor keep them per client, so
\tool seals $(\encval{\theta},\encval{m})$ into an encrypted cookie
(\S\ref{sec:cookies}). Third, the public key stores $h$ as a 32-byte seed that
a pseudorandom generator expands on demand, so the client sends the seed itself
and only $\pkval{s}$ has to be streamed as residues, which saves packets.

\section{Streaming Encapsulation}
\label{sec:keys}

The server must evaluate the ciphertext half $v=\vec{m}+\textsc{Truncate}(s\cdot
r_2+e)$ of Algorithm~\ref{alg:hqc-encaps} for a client key $s$, one packet-sized
piece at a time. The product $s\cdot r_2$ is a cyclic convolution. It is
non-local, so no contiguous block of $s$ determines any part of the output.
The CRT cannot split the ring $\Fq_2[X]/(X^n-1)$ itself into packet-sized
pieces: HQC chooses $n$ prime with 2 primitive modulo $n$, so
$X^n-1=(X-1)\,\Phi_n(X)$ with $\Phi_n$ irreducible of degree $n-1$, and the ring
has no factors of a size in between. We therefore work with the unreduced product,
which lives in $\Fq_2[X]$, where we can choose the moduli freely.

\subsection{Residues}
\label{sec:crt}

Following our overview section  (\S\ref{sec:overview}), fix public moduli
$f_1,\dots,f_k\in\Fq_2[X]$ that each have degree $\ge 1$, are pairwise coprime, and whose
degrees add up to $D\ge 2n-1$. By the CRT, the map
$w\mapsto(w\bmod f_1,\dots,w\bmod f_k)$ is a bijection from the polynomials of
degree below $D$ to the tuples of residues. Let
\[
w=\encval{\vec{m}}+\pkval{s}\cdot\encval{r_2}+\encval{e}
\]
be the \emph{unreduced} sum, computed in $\Fq_2[X]$ and not in the ring. It has
degree at most $2n-2$, so its residues fix it. Request $i$ carries the
\pkval{public-key} residue $\pkval{s_i}=\pkval{s}\bmod f_i$. The server
recomputes every \encval{encryption} term from the secret $\theta$ in the cookie
(\S\ref{sec:cookies}) and returns
\[
w\bmod f_i=\big(\pkval{s_i}\cdot(\encval{r_2}\bmod f_i)+\encval{\vec{m}}+\encval{e}\big)\bmod f_i,
\]
masked as we will describe in \S\ref{sec:mask}. The server reads a constant
amount of state and keeps nothing between requests. The client rebuilds $w$ from
the $k$ residues with Garner's algorithm~\cite{garner1959} and precomputed
constants, reduces it modulo $X^n-1$ and truncates it, which gives $v$.

\subsection{Choosing the Moduli}
\label{sec:moduli}

A $v$-residue request carries 
the sealed residue with some metadata, and all UDP payloads
are at most 1232 bytes. A residue may therefore take 1062, 1054 or 1046
bytes at HQC-1, 3 and 5, that is, at most 8496, 8432 or 8368 bits. We use the
fewest moduli that cover $2n-1$ bits under this bound, $k=5$, 9 and 14, and
split the degrees as evenly as possible, so that they add up to exactly $2n-1$
(Table~\ref{tab:moduli} in Appendix~\ref{app:moduli}). Every modulus is a trinomial $X^d+X^a+1$ with
$a\le 15$, so a reduction takes a few shifts and XORs. For each degree $d$
we computed the smallest $a$ for which the trinomial is coprime to all
earlier ones and verifies that the moduli are pairwise coprime and cover $2n-1$.

\subsection{Masking Secrets}
\label{sec:mask}

A plain HQC ciphertext is a pair of reduced ring elements $u,v$ in canonical form.
The residues pin down $w$ instead, which differs from $v$ in two ways. First, $w$ is
$v$ unreduced, giving $w = q_1(X^n-1) + v$.
Second, \textsc{Truncate} zeros bits $n_1n_2,\dots,n-1$ of $v$, but $w$ does not truncate,
giving $w = X^{n_1n_2}q_2 + v$. A naive CRT decomposition would directly reveal
$q_1$ and $q_2$, leaking intermediate algebraic values absent from a standard HQC ciphertext.
To mask each step, the server derives two masks from $\theta$ with a PRNG.
The \emph{quotient mask} $\mu$ is a random polynomial of degree at most
$n-2$. The server adds $\mu\cdot(X^n-1)$, which reduction modulo $X^n-1$
removes. The degree bound matters: $w$ has degree at most $2n-2$, so every quotient has
degree at most $n-2$ and $\mu$ covers it, while the sum keeps degree at most
$2n-2$, so the residues still fix one $w$.
The \emph{truncate mask} $\tau$ is a random polynomial of degree below
$n-n_1n_2$. The server adds $X^{n_1n_2}\tau$, which \textsc{Truncate} removes.
Reply $i$ is therefore:
\[
\big(w+\mu\cdot(X^n-1)+X^{n_1n_2}\tau\big)\bmod f_i .
\]
The masks are the same for all $k$ replies of a session, so a repeated request
gets the same reply. Write $w=q\cdot(X^n-1)+X^{n_1n_2}t+v$ with $t$ of degree
below $n-n_1n_2$. Then the masked sum equals
$(\mu+q)(X^n-1)+X^{n_1n_2}(\tau+t)+v$. The map $(\mu,\tau)\mapsto(\mu+q,\tau+t)$
is a bijection, so the masked sum is uniform on a set that depends only on $v$,
and the replies are a function of it.

\subsection{Trace Confidentiality Proof}
\label{sec:formal}

To show that the masked residues leak no secret beyond a standard HQC
ciphertext, we mechanize the masking in SSProve~\cite{ssprove} on the Rocq proof
assistant. The proof covers the algebraic masking; authentication, cookies and
the key schedule are analyzed in \S\ref{sec:security}.

We use two procedures and a simulator. $\mathtt{hqc\_encrypt}$ is HQC encryption
(Algorithm~\ref{alg:hqc-encaps}): from a public key $(h,s)$ and a message it
samples fixed-weight $(r_1,r_2,e)$ and returns $u,v$. $\mathtt{hqclethe\_encrypt}$ takes
$h$, a residue tuple $(s_1,\dots,s_k)$ and the message. If some $s$ of degree
below $n$ has exactly these residues, it samples $(r_1,r_2,e)$ as HQC does and
uniform masks $\mu$ ($n-1$ bits) and $\tau$ ($n-n_1n_2$ bits), and returns $u$
in canonical form and the $k$ masked residues of \S\ref{sec:mask}; otherwise it
fails. The simulator $\mathtt{Sim}$ gets the same input and holds no secret. It
finds the $s$ with the given residues, queries $\mathtt{hqc\_encrypt}$
with $((h,s),\vec{m})$, samples fresh $\mu',\tau'$, and returns the
resulting $u$ and the residues of $v+X^{n_1n_2}\tau'+\mu'\cdot(X^n-1)$. It never
sees $r_1$, $r_2$ or $e$.

The mechanized theorem is
$\mathtt{hqclethe\_encrypt} \;\approx_0\; \mathtt{Sim}\circ\mathtt{hqc\_encrypt}$,
SSProve's perfect indistinguishability: every distinguisher has advantage 0.
$\mathtt{Sim}$ uses the message only as input to the oracle, so the masked
residues reveal nothing beyond $(u,v)$. The moduli are parameters, with the hypotheses that they have degree $\ge 1$, are
pairwise coprime and have degrees adding up to at least $2n-1$; the development is
instantiated at HQC-1, 3 and 5. It introduces no axioms beyond the standard ones of MathComp and SSProve.
A second theorem extends this to the masks the protocol derives from $\theta$
(\S\ref{sec:conf-integrity}), showing that PRNG-derived masks remain computationally indistinguishable
from uniform. Combined with standard KDF combiner properties, a recorded session stays secret unless
both ML-KEM-768 and HQC are broken, while active protocol execution achieves server authentication
and session-key integrity under the IND-CCA security of ML-KEM-768 and AEAD integrity,
as verified symbolically in \S\ref{sec:proverif}.

\section{The \tool Protocol}
\label{sec:protocol}

\newcommand{\shared}[1]{\textcolor{blue}{#1}}
\newcommand{\phasebar}[1]{%
  \multicolumn{2}{@{}p{\linewidth}@{}}{\dotfill\ \textbf{#1}\ \dotfill}}
\newlength{\msglen}\setlength{\msglen}{4.6cm}
\newcommand{\msgrightplain}[1]{%
  \tikz[baseline=-0.5ex]\draw[thick,->] (0,0) -- (\msglen,0)
    node[midway,below=1pt]{#1};}
\newcommand{\msgrightauth}[1]{%
  \tikz[baseline=-0.5ex]\draw[thick,double,double distance=1.2pt,-{Implies}]
    (0,0) -- (\msglen,0) node[midway,below=2pt]{#1};}
\newcommand{\msgleftauth}[1]{%
  \tikz[baseline=-0.5ex]\draw[thick,double,double distance=1.2pt,-{Implies}]
    (\msglen,0) -- (0,0) node[midway,below=2pt]{#1};}

\begin{figure*}[tp]
\begin{center}
\footnotesize
\begin{tabular}{@{}p{0.47\linewidth}@{}>{\raggedleft\arraybackslash}p{0.47\linewidth}@{}}
\textbf{Client} & \textbf{Server} \emph{(\shared{blue} = shared across connections)} \\[2pt]
\phasebar{Set-up} \\[1pt]
 & static ML-KEM-768 key $(pk_{\mathrm{srv}},sk_{\mathrm{srv}})$; \shared{$s_0,s_1 \gets_\$ \{0,1\}^{256}$} \\[3pt]
\phasebar{Phase 0: Handshake} \\[1pt]
$(pk,sk)\gets_\$ \alg{HQC.KGen}$; $pk=(\mathit{seed}_h,s)$ & \\
$s_i\gets s\bmod f_i$, $i=1,\dots,k$ & \\
$(K_{\mathrm{auth}},c_{\mathrm{auth}})\gets \alg{ML\mbox{-}KEM.Encaps}(pk_{\mathrm{srv}})$ & \\
$k_{c\to s},k_{s\to c}\gets\mathrm{KDF}(K_{\mathrm{auth}})$ & \\[2pt]
\multicolumn{2}{c}{\msgrightplain{$(\mathrm{hash}(pk),\ \mathit{seed}_h,\ c_{\mathrm{auth}})$}}\\[1pt]
 & $K_{\mathrm{auth}}\gets\alg{ML\mbox{-}KEM.Decaps}(sk_{\mathrm{srv}},c_{\mathrm{auth}})$ \\
 & $m\gets_\$\{0,1\}^{k}$, $\mathrm{salt}\gets_\$\{0,1\}^{128}$, $N\gets_\$\{0,1\}^{192}$ \\
 & $(K,\theta)\gets G(\mathrm{hash}(pk),m,\mathrm{salt})$ \\
 & $K_{\mathrm{joint}}\gets\mathrm{KDF}(K,K_{\mathrm{auth}})$; $K_{\mathrm{session}},K_{\mathrm{trans}}\gets\mathrm{KDF}(K_{\mathrm{joint}})$ \\
 & $C\gets\big(N,\ \mathrm{AE}(\shared{s_b},N,\theta\|m\|\mathit{seed}_h\|K_{\mathrm{auth}}),\ b\big)$ \\
 & $\gamma\gets\mathrm{MAC}(K_{\mathrm{trans}},\,\mathrm{hash}(\mathrm{request})\|\mathrm{salt}\|b)$ \\[2pt]
\multicolumn{2}{c}{\msgleftauth{$\mathrm{AE}(k_{s\to c},\ C\|\mathrm{salt}\|\gamma,\ \mathrm{hash}(\mathrm{request}))$}}\\[3pt]
\phasebar{Phase 1: requests, independent, sent in parallel in any order} \\[1pt]
\multicolumn{2}{c}{\msgrightauth{$u$-chunk $j=0..c{-}1$: $\mathrm{AE}(k_{c\to s},\ \emptyset,\ (C, j))$}}\\[1pt]
 & $(\theta,m,\mathit{seed}_h,K_{\mathrm{auth}})\gets \mathrm{AE}^{-1}(\shared{s_b},C)$ \\
 & check the replay filter (\S\ref{sec:replay}); expand $h$ \\
 & $u_j\gets$ chunk $j$ of $r_1+h\cdot r_2$ (from $\theta$) \\[2pt]
\multicolumn{2}{c}{\msgleftauth{$\mathrm{AE}(k_{s\to c},\ u_j,\ j)$}}\\[1pt]
\multicolumn{2}{c}{\msgrightauth{$v$-residue $i=1..k$: $\mathrm{AE}(k_{c\to s},\ s_i,\ (C, i))$}}\\[1pt]
 & check, open $C$; masked residue $v_i$ modulo $f_i$ from $\theta,m$ (\S\ref{sec:mask}) \\[2pt]
\multicolumn{2}{c}{\msgleftauth{$\mathrm{AE}(k_{s\to c},\ v_i,\ i)$}}\\[3pt]
\phasebar{Phase 2: client-side recovery (local, no messages)} \\[1pt]
$u\gets u_0\|\cdots\|u_{c-1}$; rebuild from $v_1,\dots,v_k$ (CRT), reduce, truncate & \\
$K\gets \alg{HQC.Decaps}(sk,(u,v,\mathrm{salt}))$; derive $K_{\mathrm{session}}$, $K_{\mathrm{trans}}$ & \\
verify $\gamma$, else abort & \\
\end{tabular}

\vspace{3pt}
\refstepcounter{figure}\label{fig:protocol}
\small\textbf{Figure~\thefigure:} Sketch of the \tool protocol. A single arrow is a plain
datagram; the Phase-0 request is plain because clients are anonymous. A double
arrow is a datagram whose payload is sealed (authenticated and encrypted) under
a key derived from $K_{\mathrm{auth}}$.
\end{center}
\end{figure*}

This section describes the protocol. The client generates a fresh HQC key pair
and erases it after the exchange; this gives forward secrecy (\S\ref{sec:erasure}).
The server computes an ordinary HQC ciphertext $(u,v)$ for that key and never
stores more than a constant amount of memory per request.

Besides HQC and ML-KEM-768~\cite{libcrux}, \tool uses BLAKE3~\cite{blake3_github} for
hashes, MACs and key derivation (KDF), and XChaCha20-Poly1305~\cite{irtf-cfrg-xchacha-03} for
authenticated encryption (AE). 
We use ML-KEM-768 for static authentication because its 1{,}088-byte
ciphertext fits in the initial datagram. A long-term HQC key cannot take this
role on a stateless server: its ciphertext does not fit in one datagram, and
splitting HQC decapsulation across packets would hand the client residues of
the secret key.
The server derives the masks from $\theta$ with BLAKE3 and
ChaCha12 (Appendix~\ref{app:wire}).
We write $\mathrm{AE}(S,N,T,A)$ (or $\mathrm{AE}(S,T,A)$ when
the nonce is implicit) for the sealed form of plaintext $T$ under key $S$, nonce $N$, and
associated authenticated data $A$; the associated data $A$ is attached to the result
unencrypted. We write $\mathrm{AE}^{-1}$ for the verified decryption. A packet that fails
verification is dropped.

Both parties end with the same session key $K_{\mathrm{session}}$. The server
has it after Phase~0, while the client has it only after Phase~2: it must first
assemble $(u,v)$ from the replies and then run $\alg{HQC.Decaps}$ like any HQC
client. The reassembled $(u,v)$ is byte-identical to the output of the reference
$\alg{HQC.Encaps}$~\cite{hqc_impl} for the same inputs, so the client decapsulates
unmodified; our tests check this, and the reference code against the official
known-answer files, at all three levels. Only $n$ and $n_1n_2$ enter the design: if
the final standard changes $n$, the moduli are chosen again (\S\ref{sec:moduli}).

\subsection{Cookies}
\label{sec:cookies}

The server keeps two symmetric cookie keys $s_0,s_1$, shared by all
connections. A timer rotates them every 10 seconds. It draws a fresh key,
overwrites the inactive slot, and flips the active index. A cookie sealed under
$s_b$ is therefore valid for 10 to 20 seconds. A client uses its cookie only in
Phase~1, which took at most 1.6 seconds in our WAN runs. Every request carries the
one-bit key index $b$, so the server knows which key to try. If the slot has been
overwritten, the open fails and the server drops the packet. McTiny keeps a pool
of eight keys and rotates it every minute. \tool needs two, because an exchange
is far shorter.

The cookie holds everything the server needs to answer any later request:
$\theta$, $m$, the seed of $h$ and $K_{\mathrm{auth}}$. It is 152, 160 or 168
bytes at HQC-1, 3 and 5. With $K_{\mathrm{auth}}$ in the cookie, the server
derives the packet keys of each request without per-client state.

\subsection{Goals and Threat Model}
\label{sec:goals}

Our threat model assumes
the adversary controls the network: it records, drops, reorders, modifies,
replays and injects datagrams, and it may run many clients. Clients are
anonymous. The client knows the server's static ML-KEM-768 public key in advance.
We want:
\textbf{(G1)} the session key achieves secrecy and forward secrecy: recorded sessions
remain confidential against retrospective compromise of the server's static key once the
ephemeral cookie keys rotate out;
\textbf{(G2)} the client knows it talks to the holder of the static key;
\textbf{(G3)} the server keeps no per-client state: no table or allocation grows
with the number of clients, and the only state that requests change is a
fixed-size filter of request fingerprints; it never holds more than one packet, and allocates no heap after warm-up;
\textbf{(G4)} no request draws a larger reply than itself, so the server is no
amplifier;
\textbf{(G5)} all UDP payloads fit the 1232 bytes that remain of IPv6's 1280-byte minimum
MTU~\cite{rfc8200} after the IPv6 and UDP headers.
G1 provides hybrid confidentiality against passive recording: past sessions
remain secret if either ML-KEM-768 or HQC holds against a future break, with forward secrecy
against static key compromise established once the 10--20\,s cookie window elapses (\S\ref{sec:proverif}). Client
authentication and binding peer identities against unknown-key-share attacks are
left to the enclosing handshake (\S\ref{sec:conf-integrity}).
Denial of service has three attack surfaces. CPU exhaustion affects any server
that authenticates its first message; client puzzles (\S\ref{sec:puzzle})
address it. Memory exhaustion affects a server that keeps state for its
clients; G3 removes it. Saturating the network link is beyond what any protocol
can prevent and calls for filtering upstream of the server; we do not address
it.

\subsection{Phase 0: Handshake and Server Authentication}
\label{sec:handshake}

\tool authenticates the server with a static ML-KEM-768 key. The client knows the server's static
ML-KEM-768 public key. It generates an ephemeral HQC key pair
and encapsulates to the static key, which gives $K_{\mathrm{auth}}$ and a ciphertext
$c_{\mathrm{auth}}$ of 1{,}088 bytes at every HQC level. It sends $\mathrm{hash}(pk)$, the
32-byte seed of $h$ and $c_{\mathrm{auth}}$.

The server decapsulates $c_{\mathrm{auth}}$ to get $K_{\mathrm{auth}}$. It draws
$m$, a salt and a cookie nonce, and runs HQC's own function
$G(\mathrm{hash}(pk)\|m\|\mathrm{salt})$ to get $(K,\theta)$. It mixes $K$ with
$K_{\mathrm{auth}}$ into $K_{\mathrm{joint}}$, and derives from it the session key
$K_{\mathrm{session}}$ and the confirmation key $K_{\mathrm{trans}}$. The
session key therefore depends on both secrets. The server builds the cookie and
the tag $\gamma=\mathrm{MAC}(K_{\mathrm{trans}},\mathrm{hash}(\mathrm{request})\|\mathrm{salt}\|b)$,
which binds the whole request. It seals cookie, salt and $\gamma$ under
$k_{s\to c}$ (with its header and the Phase-0 request digest as associated data),
and replies.

Only the holder of $sk_{\mathrm{srv}}$ can recover $K_{\mathrm{auth}}$. A reply
that opens under $k_{s\to c}$ thus proves to the client that it talks to the
holder of the static key. The client keeps $\gamma$ and checks it at the end of
Phase~2. The server seals all session secrets statelessly into the cookie returned
to the client, maintaining zero per-connection memory and reconstructing
$K_{\mathrm{session}}$ on demand when subsequent authenticated packets arrive (\S\ref{sec:cookies}).

\subsection{Phase 1: Requests}
\label{sec:phase1}

After the handshake the client sends all requests at once. Every request
seals its payload under $k_{c\to s}$, attaching its header and the cookie
$C$ unencrypted as associated authenticated data.

\emph{$u$-chunk requests.} The client asks for each chunk of $u$, 2, 4 or 7 in
all, with an empty plaintext ($\emptyset$). The chunk index $j$ and cookie $C$ travel
unencrypted as associated data in $\mathrm{AE}(k_{c\to s},\emptyset,(C,j))$.
The server opens the cookie, expands $h$ from the seed, computes $r_1+h\cdot r_2$
with HQC's own code, and returns the bytes of the requested chunk, sealed under $k_{s\to c}$.
The server keeps nothing, so it recomputes the product for every chunk. This recomputation
is the primary CPU overhead of sending $u$ statelessly.

\emph{$v$-residue requests.} For $i=1,\dots,k$ the client sends $s_i=s\bmod f_i$ in
$\mathrm{AE}(k_{c\to s},\ s_i,\ (C, i))$. The server opens the cookie and the
request, checks that $s_i$ is reduced, computes one masked residue $v_i$ (\S\ref{sec:mask}), and returns it sealed.
A residue takes $\lceil\deg f_i/8\rceil$ bytes, at most 1030 at HQC-5. The request
is longer than the reply, because it also carries the cookie (Table~\ref{tab:packet-sizes}).

\emph{Nonces, padding and associated data.} Request nonces never go on the wire: both
endpoints derive them deterministically from the cookie nonce with its last three bytes
XORed with the direction, kind and index. One cookie, kind and index thus give one nonce.
The client builds each request once and resends the same bytes to never seal two
different plaintexts under one key and nonce. The associated data
for requests is the header and cookie $C$, so everything outside the AE payload is
authenticated as AAD, except for the request packet padding:
to prevent amplification (G4), the client pads each request with zero bytes to at least the
length of the reply it triggers. The server accepts one length per kind and rejects
non-zero padding.
Replies draw a fresh random 24-byte nonce that travels in the clear, and authenticate
the header in their AAD (plus the Phase-0 request digest in the handshake reply).

\begin{table}[t]
\caption{UDP payload per packet (request/reply, B) and packets $n$ per exchange.}
\label{tab:packet-sizes}
\centering\footnotesize\setlength{\tabcolsep}{3pt}
\begin{tabular}{@{}l rr rr rr@{}}\toprule
& \multicolumn{2}{c}{HQC-1} & \multicolumn{2}{c}{HQC-3} & \multicolumn{2}{c}{HQC-5} \\
\cmidrule(lr){2-3}\cmidrule(lr){4-5}\cmidrule(lr){6-7}
Packet kind & req/reply & $n$ & req/reply & $n$ & req/reply & $n$ \\ \midrule
Handshake & 1154/242 & 1 & 1154/250 & 1 & 1154/258 & 1 \\
$u$-chunk & 1232/1232 & 2 & 1232/1232 & 4 & 1232/1232 & 7 \\
$v$-residue & 1054/926 & 5 & 1174/1038 & 9 & 1216/1072 & 14 \\ \bottomrule
\end{tabular}
\end{table}

\subsection{Phase 2: Client-Side Recovery}
\label{sec:phase2}

The client joins the $u$-chunks. It rebuilds the masked unreduced sum from the
$k$ residues with Garner's algorithm, with precomputed constants because the
moduli are fixed. It reduces modulo $X^n-1$,
which removes the quotient mask, and truncates, which removes the truncate
mask. It then calls $\alg{HQC.Decaps}$ on $(u,v,\mathrm{salt})$ with its own $sk$,
derives $K_{\mathrm{joint}}$, $K_{\mathrm{session}}$ and $K_{\mathrm{trans}}$ as
the server did, and checks $\gamma$. On a mismatch it aborts. It erases every
intermediate value.

\subsection{Round Trips and Retransmission}
\label{sec:round-trips}

The protocol takes two sequential round trips: one for the handshake and one for
the burst of $u$-chunk and $v$-residue requests. At HQC-5 the burst has 21
packets and 25{,}648 bytes. McTiny needs about eleven round trips and 971 packets
in each direction. Its row-wise combination cannot start before all 952
first-phase replies are in, and its key stream is paced by client-side
congestion control.

The client keeps one retransmission timer per phase. It starts at 500\,ms for
the handshake, and at three times the measured handshake round-trip time for the
burst; both are bounded to the range 100\,ms to 2\,s. The timer doubles at every stall and the client
gives up after six stalls in a row in the handshake and seven in the burst. After a
stall it resends only the requests that have no reply yet.
The burst uses no congestion control. At HQC-5 it is $1.7\times$ the 14{,}720-byte
initial window of QUIC and TCP~\cite{rfc9002,rfc6928}, and on a congested path it can
lose several packets at once. A deployment that must stay within the window sends
the burst in two windows, one round trip apart. No reply is longer than its request
(Table~\ref{tab:packet-sizes}), so the burst amplifies nothing.

\subsection{Replay Filter}
\label{sec:replay}

A stateless server cannot tell a replayed request from the first copy, so anyone
who sees a sealed Phase-1 request could exhaust the server's CPU by making it redo the work while the
cookie is valid. After the cookie and AE checks,
the server hashes (cookie nonce, kind, index) under a per-process key
and looks the fingerprint up in a Bloom filter. A new fingerprint is inserted and
answered; a known one is answered at most 3 more times to let in honest
retransmissions but not malicious spam. The fingerprint holds nothing else, so the replay limit
applies uniquely to the packet type and the corresponding session, but not uniquely to the packet content itself.
The filter keeps two generations and starts a new one at each
cookie-key rotation. An entry therefore lives until its cookie can no longer be
opened. Filter size is constant and calculated from
the server's compute bandwidth and a target false-positive rate of 0.1\,\%
(Appendix~\ref{app:replay}); this is 574 to 1175\,KiB per worker on our x86 host
and 30 to 90\,KiB on the Pi. The filter stores fingerprints of client requests, but it is fixed-size,
independent of the number of clients, and allocated before the first request, so
it keeps G3.

\subsection{Client Puzzle}
\label{sec:puzzle}

A handshake request carries no cookie, so the replay filter cannot limit it, and
each one costs the server an ML-KEM-768 decapsulation. An
attacker can therefore keep the server busy with new handshake requests. As in
HIP~\cite{rfc7401}, IKEv2~\cite{rfc8019} and Tor onion services~\cite{tor-prop327},
the server can then make the client do work first. The operator turns this client
puzzle on and sets its difficulty $d$ at run time. While it is on, the server
answers a handshake request with a 55-byte challenge instead of decapsulating. The
challenge holds a nonce, $d$, a time and a MAC over these and the client's IP
address, under a key derived from the cookie key. The client searches for a counter
such that the BLAKE3 hash of the challenge, its address, a 32-byte digest of its
request and the counter starts with $d$ zero bits, and resends the request with the
challenge and the counter.
The server checks the MAC, the time and the hash, admits the challenge through the
replay filter (\S\ref{sec:replay}), and only then decapsulates. It derives $m$, the
salt and the cookie nonce from the challenge and the request packet with a PRF whose
key is derived from the cookie key, so one solved puzzle only buys one session, and
the PRF key is erased with the cookie key within 10--20\,s, which keeps forward
secrecy (G1) in puzzle mode. The replay filter answers a repeat of the
solved request at most three more times, so it buys at most four decapsulations,
all for that session. The puzzle adds one round trip but only while it is on.


\section{Security Analysis}
\label{sec:security}

We analyze the protocol's security both in prose and formally. \S\ref{sec:formal} proves the
masking mechanism, and ProVerif checks the protocol logic (\S\ref{sec:proverif}). We use the goals and the network adversary of \S\ref{sec:goals}.

\subsection{Confidentiality and Integrity}
\label{sec:conf-integrity}

\emph{What protects each packet.} The Phase-0 request is plain. It holds
$\mathrm{hash}(pk)$, the seed of $h$, and $c_{\mathrm{auth}}$; none of these is
secret as per G1, and $c_{\mathrm{auth}}$ is an ML-KEM-768 ciphertext. The
cookie is sealed under the server's cookie key. All other payloads are sealed
under keys derived from $K_{\mathrm{auth}}$.

\emph{Passive attackers.} An eavesdropper observes the cookie, the $c$ chunks of $u$,
and the $k$ residues. Confidentiality rests on three properties:
(1)~while $K_{\mathrm{auth}}$ is secret, the sealed packets reveal nothing
beyond their length. (2)~If $K_{\mathrm{auth}}$ leaks later, because the static
ML-KEM-768 key leaks, the attacker reads the plain payloads. These are the public
key $pk$, the chunks of $u$, and the masked residues. By the theorem of
\S\ref{sec:formal}, this reveals nothing beyond a pair of plain HQC ciphertext
and the corresponding public key.
(3)~The session key is
$\mathrm{KDF}(K,K_{\mathrm{auth}})$ and so is secret if the HQC key $K$ is
secret, which follows from the IND-CCA security of HQC.

\emph{Active attackers.} On the server's end, a forged cookie fails verification, and a
request without $k_{c\to s}$ fails the AE check; the server learns nothing from a failed
request and keeps nothing. The associated data of a request covers the header and the
cookie, so a cookie cannot be moved to another header, and a third party that copies a
cookie cannot seal a request unless it holds the static key (\S\ref{sec:proverif}).
On the client's end, a forged or changed reply fails the AE
check, which needs $k_{s\to c}$ and so $K_{\mathrm{auth}}$. The tag $\gamma$ confirms that
the client's decapsulated $K$ equals the server's and binds the whole Phase-0 request;
corrupted residues make decapsulation return a pseudorandom key that fails $\gamma$.
An attacker that poses as the server cannot open $c_{\mathrm{auth}}$ without
the static secret key, so it cannot derive $k_{s\to c}$ or produce a reply that
the client accepts. An attacker that holds the static key can pose as the
server, but so can such an attacker in any protocol, since without some key
that both sides trust in advance, no client can tell the real server from an
impostor.

\emph{Malicious clients.} A client that completed Phase~0 can send any residue, even
one not from a valid key; chosen residues reveal the secrets $r_2$, $e$ and $m$,
which is meant for the client anyway. Cookies are independent across sessions, so it cannot
access secrets meant for other clients.
The server binds $K$ to $H(pk)$ but does not check that the
residues belong to $pk$; a client that streams another key only fails to decapsulate. In actual deployments, a
full handshake should also bind the peers' identities against unknown-key-share attacks;
$\gamma$ binds only the Phase-0 request.

\emph{Security against quantum computers.}\label{sec:quantum} 
The protocol rests on HQC and ML-KEM's IND-CCA security~\cite{fujisaki_okamoto,fo_modular}.
This is a safe assumption given NIST has selected both for standardization~\cite{nist-hqc-selection,alagic2024module}.
Authentication rests on ML-KEM alone, which suffices because it only has to
hold while a session is set up: an attacker who breaks ML-KEM later cannot go
back and impersonate the server. Confidentiality must hold for as long as the
recorded session matters, which is why the session key also depends on HQC.
One minor argument against HQC is that besides the hardness assumption, the security argument also
rests on its decryption failure rate (DFR).
A ciphertext has a small probability of failing to be decrypted, and failures
leak the secret key~\cite{hqc_dfr_attack}.  While the DFR is argued to be
negligibly small by models~\cite{hqc,bdmp25,mtv25,dfr26}, no unconditional proof
exists yet, but \tool does not reuse HQC public keys for more than one session, so a failure would only abort not compromise
the session, assuming existing works in fact overestimate the DFR.
The symmetric primitives in \tool use
256-bit keys, which Grover's algorithm~\cite{grover} reduces only to 128 bits.

\subsection{Reflection and Amplification}
\label{sec:amplification}

Phase~0 has an unauthenticated request and a source address that may be forged.
Padding makes every request at least as long as the reply it triggers. At every
level the handshake request is longer than its reply (1{,}154 against 242, 250
and 258 bytes). For $u$-chunks the
ratio is 1.00 for all but the shorter last chunk, and for $v$-residues it is
0.88. Reply size
never exceeds request size, so the server is no amplifier (G4). QUIC uses the same remedy
with its 1200-byte floor for initial packets~\cite{rfc9000}. Padding costs 13 to
15\,\% of the bytes of a session (Table~\ref{tab:bandwidth}). A third party that
captures a sealed request can resend it with a forged source address; the
replay filter (\S\ref{sec:replay}) bounds the number of times the server would
reply to the forged address, limiting the damage of reflection attacks.

\subsection{Key Erasure and Forward Secrecy}
\label{sec:erasure}

Forward secrecy (G1) asks what an attacker can recover who records a session now and
later compromises either endpoints. The client creates a new key pair for each
connection and wipes the secret key, the key-generation entropy, $K$ and
$K_{\mathrm{trans}}$ at the end of Phase~2 or on any abort. The server wipes
$\theta$, $m$ and the derived keys when each handler returns. What remains on
the server is the static ML-KEM-768 key and the two cookie keys. The cookie keys
are overwritten within 10--20 seconds by a timer that no party can postpone.

After that time, the server holds nothing from which $K_{\mathrm{session}}$ can
be rebuilt. An attacker who steals the server and has recorded the exchange
can recover $K_{\mathrm{auth}}$, and with it read all payloads, but then faces
the masked residues, which carry no more than a plain HQC ciphertext, and
$K$, which needs the erased client key. For connections younger than 20
seconds the attacker can reopen the cookie and compute the session key from
it, as in McTiny~\cite[Section~7.1]{mctiny}. We improve on this in comparison
since we make very short-lived sub-minute cookie keys practical.

\begin{table}[t]
\caption{Summary of formal verification results across Rocq/SSProve and ProVerif.}
\label{tab:proved}
\centering\scriptsize\setlength{\tabcolsep}{2pt}
\begin{tabular}{@{}>{\raggedright\arraybackslash}p{0.34\columnwidth}>{\raggedright\arraybackslash}p{0.18\columnwidth}>{\raggedright\arraybackslash}p{0.44\columnwidth}@{}}\toprule
Claim & Framework & Assumptions; Scope \\ \midrule
Masked replies reveal only $(u,v)$ & Rocq/SSProve & Uniform masks; perfect simulation \\
Same, with masks derived from $\theta$ & Rocq/SSProve & PRNG mask stream; computational \\
Session secrecy \& forward secrecy & ProVerif & Unbounded sessions; hybrid KEM \\
Server authentication & ProVerif & Static ML-KEM-768 key; transcript agreement \\
Replay immunity & ProVerif & Once-only filter; bounded answers \\
Necessity of server checks & ProVerif & Attack trace found if any check removed \\
Puzzle binding & ProVerif & Also with later compromise of the static or cookie key \\
\bottomrule\end{tabular}\end{table}

\subsection{Symbolic Analysis of the Protocol}
\label{sec:proverif}

Rocq/SSProve proves the simulatability of the residue stream
(\S\ref{sec:formal}); ProVerif~2.05~\cite{proverif} checks full protocol
execution and active attacks. The model uses ideal ML-KEM-768 and HQC KEMs
(with the HQC ciphertext sent in pieces), an ideal AEAD, and random-oracle
hashes.
ProVerif cannot model polynomial arithmetic, so the model treats a reply to a
residue from the honest client as revealing nothing, which \S\ref{sec:formal}
justifies, and assumes that a reply to any other residue
reveals the session seed~$m$, the worst case. As is common, the two
analyses are separate: SSProve proves the masking computationally, and ProVerif
analyses the protocol in the symbolic model, taking that result as an
assumption.

The proofs support the arguments made in \S\ref{sec:conf-integrity} and \S\ref{sec:erasure}.
For unbounded sessions, ProVerif proves secrecy of the session key and $m$,
transcript agreement, and once-only request handling, including for
attacker-chosen cookies. With the puzzle enabled, the server decapsulates
only after issuing a challenge to the requester's address and receiving a
valid solution for that request; each challenge is used at most once.
Removing any server check yields an attack: requests can be replayed,
decapsulation obtained without solving the puzzle for the bound address, or
a residue answered without a corresponding client request.
Forward secrecy still holds after ephemeral cookie keys rotate out, at most
20\,s after session initiation, even if the static ML-KEM-768 key is later
compromised. A later ML-KEM-768 break may recover $c_{\mathrm{auth}}$, but
does not recover the HQC session key because the client has erased its
ephemeral private key. As explained above, an attacker who already holds the static key during
the handshake can impersonate the server, which is outside our scope.
\section{Implementation and Evaluation}
\label{sec:evaluation}

We ask:
\textbf{(Q1)} does \tool make a stateless code-based hedge practical compared
with McTiny (\S\ref{sec:eval-hedge});
\textbf{(Q2)} what does statelessness cost relative to a stateful server (\S\ref{sec:eval-price});
\textbf{(Q3)} does statelessness hold up under attack, against both CPU floods and memory exhaustion (\S\ref{sec:eval-dos});
\textbf{(Q4)} do the results hold on low-power hardware (\S\ref{sec:eval-arm}).

\tool is written in Rust. HQC key generation, decapsulation, sampling and the
product $r_1+h\cdot r_2$ come from the official reference
implementation~\cite{hqc_impl}, so the endpoints are byte-compatible with plain
HQC. New Rust code covers the residues, the masks, Garner's algorithm,
cookies, the replay filter, the client puzzle, packet parsing and the server loop. On x86 we use the AVX2 build of HQC and of
libcrux ML-KEM~\cite{libcrux}. Multiplication modulo a trinomial uses carry-less
multiplication. 
A Valgrind check in the style of ctgrind finds no
secret-dependent branch or address in the server code for $u$ and $v$. \ifarxiv An artifact\else An anonymized artifact\fi{} with the code and the
proof is at:
\begin{center}
\artifacturl
\end{center}

\subsection{Comparison Points and Setup}
\label{sec:setup}

We measure the whole ephemeral code-based key exchange: a fresh ephemeral code-based key pair, its
encapsulation streamed to a stateless server, and the derived key. Only a protocol that
also streams a code-based key in pieces to a server without per-client state is
comparable, and McTiny is the only published one; \tool builds on its design.
Rosenpass and PQ-WireGuard~\cite{rosenpass,pq-wireguard} finish in one round trip, but
only because they do not use Classic McEliece ephemerally, so they are incomparable. We run the reference release \texttt{mctiny-20200804} with
SUPERCOP 20200702 and \texttt{mceliece6960119}, unmodified, on the same host. This matches
the security level of HQC-5: both are NIST category~5.
McTiny uses Classic McEliece for server authentication, \tool ML-KEM-768. We also report HQC-1 and
HQC-3. They target lower security categories than McTiny's key, and are
therefore not directly comparable.

To separate the cost of statelessness from the cost of HQC, we
compare with the strongest simple alternative, a
server that buffers the key after a cookie. It uses the same Phase~0, the same
keys from $K_{\mathrm{auth}}$ and the same sealed packets. The client sends a handshake request with $c_{\mathrm{auth}}$. The
server decapsulates, and returns a cookie that is bound to the client's address
and valid for 30 seconds. The client uploads its HQC public key in 2, 4 or 7
sealed fragments. A table entry is created only after a valid cookie and a valid fragment. When the key is
complete, the server runs ordinary HQC encapsulation and returns the ciphertext
in sealed fragments. The table has a cap of 1024 entries by default, at most 4
entries per source /64 prefix, an idle timeout of 3 s, a lifetime of at most 10 s,
and evicts only entries older than 1 s. An entry costs the key size plus 160
bytes: 2.4, 4.7 and 7.4\,KB at HQC-1, 3 and 5.

All x86 results come from one AMD Ryzen 7 5800H with AVX2 at a fixed 3169\,MHz,
with the server on its own core; Appendix~\ref{app:x86} gives the testbed details.

\subsection{Practical Code-Based Hedging}
\label{sec:eval-hedge}
\label{sec:eval-network}
\label{sec:eval-cpu}
\label{sec:eval-latency}

\begin{table}[t]
\caption{Bytes per key exchange (UDP payload). McTiny uses \texttt{mceliece6960119}.}
\label{tab:bandwidth}
\centering\footnotesize\setlength{\tabcolsep}{3pt}
\begin{tabular}{@{}lrrrr@{}}\toprule
 & HQC-1 & HQC-3 & HQC-5 & McTiny \\ \midrule
Packets, \tool & 16 & 28 & 44 & 1{,}942 \\
Packets, stateful & 8 & 14 & 22 &  \\
Upstream, \tool & 8{,}888 & 16{,}648 & 26{,}802 & 1{,}188{,}422 \\
Downstream, \tool & 7{,}165 & 14{,}242 & 22{,}765 & 135{,}977 \\
Total, \tool & 16{,}053 & 30{,}890 & 49{,}567 & 1{,}324{,}399 \\
Total, stateful & 8{,}317 & 15{,}531 & 24{,}241 &  \\
Raw key + ciphertext & 6{,}674 & 13{,}492 & 21{,}658 & 1{,}047{,}545 \\
\tool total / raw & 2.41$\times$ & 2.29$\times$ & 2.29$\times$ &  \\
\tool padding & 2{,}124 & 4{,}216 & 7{,}322 &  \\
Round trips & 2 & 2 & 2 & 11--12 \\
\bottomrule\end{tabular}\end{table}

We answer \textbf{(Q1)} affirmatively. At category~5, \tool sends $27\times$ fewer bytes than McTiny,
costs the server $9.5\times$ less CPU, and finishes about $6\times$ sooner;
under loss the gap grows to minutes.

\tool needs 16, 28 and 44 packets and 16.1, 30.9 and 49.6\,KB at
HQC-1, 3 and 5 (Table~\ref{tab:bandwidth}). At category~5, McTiny needs 1{,}942
packets and 1{,}324{,}399 bytes, so \tool sends $44\times$ fewer packets and
$27\times$ fewer bytes. This saving stems from key size: the HQC-5 key is
$145\times$ smaller than McTiny's. Our streaming itself costs more per byte:
\tool spends 2.3--2.4 bytes per raw byte, against McTiny's 1.26;
13--15\,\% is no-amplification padding, and residues carry the unreduced sum (14
replies of 1030 data bytes for 7.2\,KB of $v$, a factor of 2.0). At HQC-5,
33\,\% of the traffic belongs to $u$-chunks and $v$-residues 65\,\%.

\begin{table}[t]
\caption{Server CPU (Ryzen 7 5800H, 3169\,MHz). Overhead: (session CPU $-$ handler costs) / requests per session.}
\label{tab:cpu}
\centering\small\setlength{\tabcolsep}{4pt}
\begin{tabular}{@{}llrrr@{}}\toprule
 & & HQC-1 & HQC-3 & HQC-5 \\ \midrule
Handshake (\textmu s) & & 32.6 & 33.1 & 33.1 \\
$u$-chunk (\textmu s) & & 34.6 & 65.5 & 124.1 \\
$v$-residue (\textmu s) & & 31.1 & 45.6 & 64.4 \\
Overhead per request (\textmu s) & & 16.2 & 13.8 & 12.2 \\
\midrule
Session, \tool (ms) & & 0.387 & 0.899 & 2.071 \\
Session, stateful (ms) & & 0.207 & 0.312 & 0.507 \\
McTiny / \tool & & --- & --- & 9.5$\times$ \\
\bottomrule\end{tabular}\end{table}

We measure the CPU time of the server process per session with Linux
\texttt{schedstat}, summed over all server threads, before and after 1000
sessions, in five rounds and two independent runs (100 sessions per round for
McTiny). The figure includes the kernel time spent in the server's system calls
and excludes the client. No session of \tool or the baseline failed (60{,}000 in total). Four of 1{,}500 McTiny sessions
timed out, likely due to a lost datagram on the loopback; their CPU is divided by the completed
sessions, and the mean decreases from 19.67\,ms to 19.61\,ms if rounds affected by timeouts are excluded.
One \tool session costs the server 0.39, 0.90, 2.07\,ms at categories~1,3,5 respectively,
whereas McTiny costs 19.67\,ms at category 5, a ratio of $9.5\times$ (Table~\ref{tab:cpu}),
close to the 53 million cycles that its authors report \cite{mctiny}.
McTiny handles 1{,}942 packets per session and \tool 44.
We also measure \tool with Criterion~\cite{criterion}, which measures the server request handlers
without system calls, replay filtering, and kernel networking. This gives
0.25, 0.70, 1.80\,ms at categories~1,3,5 respectively.
Handshakes are dominated by ML-KEM-768 decapsulation
(24.8\,\textmu s of 33.1\,\textmu s at HQC-5); $u$-chunks are the costliest
requests (124\,\textmu s), requiring the full product $h\cdot r_2$. Computing only
the requested chunk could be cheaper, because $r_2$ is sparse, but the positions
of its nonzero coefficients are secret, so a constant-time version needs
rotations by secret amounts; we have not built one and leave it to future work.

\begin{figure}[t]
\centering
\begin{tikzpicture}
\begin{semilogyaxis}[evalplot, width=\columnwidth, height=4.6cm,
  xlabel={Packet loss per direction (\%), 20\,ms RTT}, ylabel={Session time (s)},
  xmin=-0.3, xmax=5.3, ymin=0.02, ymax=300, xtick={0,1,2,2.5,3,4,5},
  xticklabels={0,1,2,2.5,3,4,5}, ytick={0.1,1,10,100}, yticklabels={0.1,1,10,100},
  legend pos=north west]
\addplot+[hqclethe, error bars/.cd, y dir=both, y explicit] coordinates {(-0.07,0.0449) -= (0,0.0033) += (0,0.0031) (0.93,0.0494) -= (0,0.0047) += (0,0.2311) (1.93,0.1649) -= (0,0.1223) += (0,0.2653) (2.43,0.1065) -= (0,0.0644) += (0,0.0651) (2.93,0.1642) -= (0,0.1203) += (0,0.0090) (3.93,0.1678) -= (0,0.1241) += (0,0.2382) (4.93,0.1692) -= (0,0.0034) += (0,0.1671)};
\addplot+[stateful, error bars/.cd, y dir=both, y explicit] coordinates {(0.00,0.0459) -= (0,0.0043) += (0,0.0023) (1.00,0.0467) -= (0,0.0038) += (0,0.2004) (2.00,0.0483) -= (0,0.0042) += (0,0.4936) (2.50,0.0468) -= (0,0.0052) += (0,1.1750) (3.00,0.1481) -= (0,0.1048) += (0,1.0367) (4.00,0.1444) -= (0,0.1015) += (0,0.6806) (5.00,0.2757) -= (0,0.2312) += (0,1.2131)};
\addplot+[mctiny, error bars/.cd, y dir=both, y explicit] coordinates {(0.07,0.2791) -= (0,0.0260) += (0,0.0294) (1.07,1.1172) -= (0,0.4182) += (0,1.7653) (2.07,3.6881) -= (0,2.4388) += (0,14.2818) (2.57,7.1392) -= (0,4.8788) += (0,113.2813) (3.07,51.6505) -= (0,49.5147) += (0,69.0589) (4.07,103.1275) -= (0,89.7021) += (0,18.5774) (5.07,120.6853) -= (0,101.1175) += (0,7.4198)};
\legend{\tool, Stateful, McTiny}
\end{semilogyaxis}
\end{tikzpicture}

\caption{Session time at category~5 under uniform loss in both directions (20\,ms RTT): median, with whiskers from the 5th to the 95th percentile (30 sessions per point).}
\label{fig:hedge}
\end{figure}
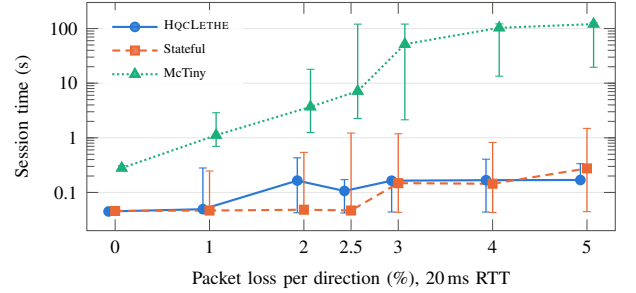

We time
\tool from its first request to the derived key, on five wide-area profiles, a loss
sweep at 20\,ms RTT with uniform loss in both directions, and a bursty
Gilbert--Elliott process of about 5\,\% loss~\cite{gilbert1960capacity,elliott1963estimates}, with 30 sessions per cell (Table~\ref{tab:wan} in Appendix~\ref{app:tables}).
For McTiny we use the McTiny client's own packet log and time from the first packet sent to the last reply received.
Before any network traffic, the McTiny client spends 0.06 to 0.11\,s per session,
mostly on a fresh Classic McEliece key pair; \tool spends 4 to 7\,ms.
Without loss, \tool needs two round trips: 0.166\,s at 75\,ms and 0.636\,s at
300\,ms. McTiny needs about 11 to 12 (its authors report 10.9~\cite{mctiny}) and is
$5.9\times$ to $6.0\times$ slower from 75\,ms upward.
Loss widens the gap (Fig.~\ref{fig:hedge}). A session that loses a packet waits one retransmission timer of at least
0.1\,s. \tool's median is 0.049\,s at 1\,\% and 0.11 to 0.17\,s from 2 to 5\,\%, and
its 95th percentile in the loss sweep is at most 0.43\,s. McTiny's median is 3.7\,s at 2\,\% and about 121\,s at 5\,\%, largely
because its retransmission timer doubles up to 120\,s.

\subsection{The Price of Statelessness}
\label{sec:eval-price}

\begin{table}[t]
\caption{Server memory footprint and binary size.}
\label{tab:memory}
\centering\small\setlength{\tabcolsep}{4pt}
\begin{tabular}{@{}lrrrr@{}}\toprule
 & HQC-1 & HQC-3 & HQC-5 & McTiny \\ \midrule
Stack, \tool (KiB) & 100 & 100 & 104 & 84 \\
Stack, stateful (KiB) & 44 & 44 & 44 & \\
RSS, \tool (MB) & 4.1 & 3.9 & 3.8 & 2.4 \\
RSS, stateful (MB) & 3.0 & 3.1 & 3.1 & \\
Binary, \tool (KB) & \multicolumn{3}{c}{1142} & 302 \\
Binary, stateful (KB) & \multicolumn{3}{c}{1491} & \\
\bottomrule\end{tabular}\end{table}

We quantify \textbf{(Q2)} to about $2\times$ the bytes and $2$--$4\times$ the server CPU, in
exchange for a fixed 1.2--1.6\,MiB memory footprint. A stateless server cannot keep the key it has
received, so it repeats work and resends data instead. In return, its memory
no longer depends on its clients, so an attacker cannot exhaust it.

Across all 3 security levels, the stateful baseline sends 1.12 to 1.25 bytes per
raw byte: 8.3, 15.5 and 24.2\,KB in 8, 14 and 22 packets. \tool costs 16.0,
30.9, 49.6\,KB instead. Both take two sequential round trips, but under loss
the stateful server is often faster: at 2\,\% loss and under Gilbert--Elliott
loss its median is 0.048 and 0.049\,s, against 0.165 and 0.170\,s for \tool
(Fig.~\ref{fig:hedge}). \tool sends twice as many packets, so a session is more
likely to lose one and wait for a retransmission timer of at least 0.1\,s.
The stateful baseline needs 0.21, 0.31 and 0.51\,ms of
server CPU per session. \tool costs $1.9\times$, $2.9\times$ and $4.1\times$ as
much. The difference is the work that the stateless server repeats: it
recomputes $u$ for every chunk and $r_2$, $e$ and the masks for every residue,
while the baseline encapsulates once. We note the price is still small: at
HQC-5 a session only costs 50\,KB and 2.1\,ms, which is negligible in terms of real application
traffic and latency.

In exchange we obtain fixed memory. When measuring the minimum stack limit
(\texttt{ulimit -s} in 4\,KiB steps) under which a full session succeeds, \tool
needs 100 to 104\,KiB of stack, $1.2\times$ McTiny's 84\,KiB at HQC-5, and a
binary of 1.14\,MB, $3.8\times$ McTiny's 302\,KB (Table~\ref{tab:memory}). The
stateful baseline needs 44\,KiB of stack at every level and a 1.49\,MB binary,
mostly due to code Rust libraries pulls in.
Resident memory is 3.3 to 3.4\,MB idle and 3.7 to 4.1\,MB after 20 sessions, as
sessions fill the replay filter of 0.6 to 1.2\,MB; the stateful baseline uses
2.8 to 2.9\,MB idle and 3.0 to 3.1\,MB after 20 sessions.
We also ran each server on the Raspberry Pi (\S\ref{sec:eval-arm}) inside a
Linux control group with no swap. We searched in 64\,KiB steps for the smallest
memory limit under which every session completes and the kernel does not re-read
the server binary from storage. The loads were one session, 50 sessions in
sequence, and 16 or 64 concurrent clients.
\tool makes 26 \texttt{brk}/\texttt{mmap} calls at start-up and none afterwards,
it only needs 1.2 to 1.6\,MiB at every level and every load. Its anonymous memory
is the same for 1 and 64 clients. The stateful baseline grows by one measured
key buffer for each pending client: 2{,}401, 4{,}674 and 7{,}397\,B at HQC-1, 3
and 5. With 64 clients and 20\,\% loss it held 41 pending clients at HQC-5, and
its anonymous memory grew by about 320\,KiB. McTiny needs about 0.5\,MiB,
because its binary is smaller. Most of the limit is binary code that must stay
resident; the working memory of \tool is 0.3 to 0.4\,MiB. A stateful server must
instead reserve memory in proportion to the cookies an attacker can hold
(\S\ref{sec:eval-dos}).

One side advantage with \tool is that we get free replication.
A buffering server must see every fragment of one upload, so
replicas behind anycast or a load balancer need session affinity or a shared table, and
an anycast route change or NAT rebinding mid-upload restarts the upload. In contrast, \tool's
replicas share only the static key and the cookie keys, so any of them can answer any
packet; a one-way hash ratchet could allow agreement on
cookie keys without a synchronization channel while keeping erased keys unrecoverable.
Each replica keeps its own replay filter, so an attacker that sends the same
requests to $R$ replicas gets up to $R$ times the answers. Behind anycast, an
attacker reaches only its nearest replica; behind a load balancer that spreads
one client over all replicas, the operator raises $d$ by $\log_2 R$ to keep the
same work per puzzle.

\subsection{Under Attack}
\label{sec:eval-dos}

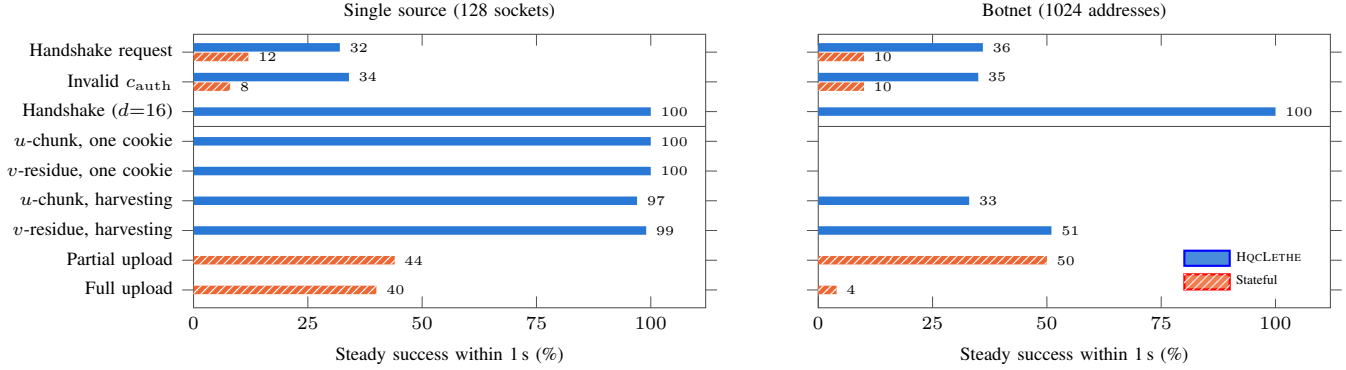
\begin{figure*}[t]
\centering
\pgfplotsset{floodpanel/.style={evalplot, xbar=0pt, bar width=3.2pt, width=0.47\textwidth, height=5.2cm,
  xmin=0, xmax=112, xtick={0,25,50,75,100}, ytick={0,...,8}, yticklabels={{Handshake request},{Invalid $c_{\mathrm{auth}}$},{Handshake ($d{=}16$)},{$u$-chunk, one cookie},{$v$-residue, one cookie},{$u$-chunk, harvesting},{$v$-residue, harvesting},{Partial upload},{Full upload}}, y dir=reverse,
  ymin=-0.6, ymax=8.6, ymajorgrids=false, xlabel={Steady success within 1\,s (\%)},
  nodes near coords, nodes near coords style={font=\tiny, anchor=west, text=black}}}
\begin{tikzpicture}
\begin{axis}[floodpanel, title={Single source (128 sockets)}]
\addplot+[hqclethebar, bar shift=1.8pt] coordinates {(32,0) (34,1)};
\addplot+[hqclethebar, bar shift=0pt, forget plot] coordinates {(100,2) (100,3) (100,4) (97,5) (99,6)};
\addplot+[statefulbar, bar shift=-1.8pt] coordinates {(12,0) (8,1)};
\addplot+[statefulbar, bar shift=0pt, forget plot] coordinates {(44,7) (40,8)};
\draw[floodsplit] (axis cs:0,2.5) -- (axis cs:112,2.5);
\end{axis}
\end{tikzpicture}\hfill
\begin{tikzpicture}
\begin{axis}[floodpanel, title={Botnet (1024 addresses)}, yticklabels={}, legend pos=south east]
\addplot+[hqclethebar, bar shift=1.8pt] coordinates {(36,0) (35,1)};
\addplot+[hqclethebar, bar shift=0pt, forget plot] coordinates {(100,2) (33,5) (51,6)};
\addplot+[statefulbar, bar shift=-1.8pt] coordinates {(10,0) (10,1)};
\addplot+[statefulbar, bar shift=0pt, forget plot] coordinates {(50,7) (4,8)};
\legend{\tool, Stateful}
\draw[floodsplit] (axis cs:0,2.5) -- (axis cs:112,2.5);
\end{axis}
\end{tikzpicture}
\caption{Legitimate success during floods at HQC-5, one server core. Floods below the line first obtain a cookie. A missing bar means the flood does not apply or was not run.}
\label{fig:flood}
\end{figure*}

\begin{figure}[t]
\centering
\begin{tikzpicture}
\begin{loglogaxis}[evalplot, width=\columnwidth, height=4.8cm,
  xlabel={Handshake requests the attacker gets answered (per s)}, ylabel={Server state (MB)},
  xmin=10, xmax=100000, ymin=0.1, ymax=5000, clip=false,
  ytick={0.1,1,10,100,1000}, yticklabels={0.1,1,10,100,1000},
  xtick={10,100,1000,10000,100000}, xticklabels={10,100,1k,10k,100k}]
\addplot[hqcletheband] coordinates {(10,1.26) (100000,1.26) (100000,1.68) (10,1.68)} -- cycle;
\node[hqclethelabel] at (axis cs:2000,1.26) {\tool, any rate: 1.2--1.6\,MiB};
\addplot[pi ram, domain=10:100000] {937};
\node[ramlabel] at (axis cs:12,937) {Pi 3B+ RAM};
\addplot[statefulone, domain=10:100000] {x*10*2401/1e6};
\node[statelabel] at (axis cs:100000,2401.0) {HQC-1};
\addplot[statefulthree, domain=10:100000] {x*10*4674/1e6};
\node[statelabel] at (axis cs:100000,4674.0) {HQC-3};
\addplot[statefulfive, domain=10:100000] {x*10*7397/1e6};
\node[statelabel] at (axis cs:100000,7397.0) {HQC-5};
\addplot[measured] coordinates {(115,2.76) (115,5.38) (115,8.51)};
\node[ratelabel, anchor=north west] at (axis cs:115,2.76) {1 attacker core, $d{=}16$};
\addplot[measured] coordinates {(11075,265.91) (11075,517.65) (11075,819.22)};
\node[ratelabel, anchor=south east] at (axis cs:11075,819.22) {4 Pi cores, no puzzle};
\end{loglogaxis}
\end{tikzpicture}
\caption{Memory an attacker can make the stateful server hold once its cookies are shortened from 30\,s to \tool's 10\,s. Dots: measured handshake request rates. Band: \tool's fixed memory.}
\label{fig:state}
\end{figure}
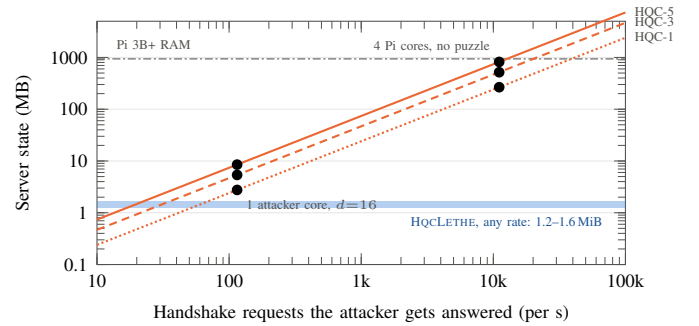

We answer \textbf{(Q3)} affirmatively, taking the three attack surfaces of
\S\ref{sec:goals} in turn. CPU exhaustion hurts both designs: handshake floods
are the strongest attack on \tool, but they hurt the stateful server even more,
and the puzzle mitigates them; floods of $u$-chunk and $v$-residue requests,
each of which recomputes part of the HQC encapsulation, are capped by the replay
filter (Fig.~\ref{fig:flood}). Memory exhaustion works only against the
stateful server (Fig.~\ref{fig:state}), whereas \tool is immune. Saturating the
network link shows up on the Pi (\S\ref{sec:eval-arm}).

To measure the impact of exhaustion attacks, we flood one server core for 10\,s
and start a legitimate probe every 200\,ms, from one source (128 sockets) or
from a botnet of 1024 addresses. \emph{Steady success} is the share of probes
started during the flood that finish within 1\,s (median of 10 repetitions; 5
for harvesting floods). Appendix~\ref{app:tables} describes the flood kinds and
gives all results. Every handshake request forces an
ML-KEM-768 decapsulation, and the replay filter cannot help, because a
handshake request carries no cookie yet. The server core is saturated and
answers about 26{,}000 handshake requests per second at every level; steady
success falls to 52, 42 and 32\,\% at HQC-1, 3 and 5. The same flood leaves
the stateful server only 12 to 23\,\%.

The client puzzle works as a mitigation. With the puzzle on ($d=16$), the
server answers handshake requests with 55-byte challenges instead of
decapsulations. It answers all of about 60{,}000 requests per second at 68 to
69\,\% CPU, and steady success is 100\,\% at every level. The request is
1{,}154 bytes, so the puzzle does not amplify.
An attacker that solves puzzles still gets decapsulations, but pays for them.
The hashed input is 111 bytes, so every attempt costs one BLAKE3 compression;
the honest solver already reuses the fixed prefix, so an attacker gains nothing
by caching it.
At $d=16$ a solution costs about 8.7\,ms on one x86 core, and buys at most four
sessions' worth of server work per worker, because the replay filter answers
each request of its cookie at most four times: 8.3\,ms at HQC-5 on x86 and
126\,ms on the Pi. One attacker core therefore keeps about one x86 worker
busy, but saturates a Pi worker with 0.07 of a core: at $d=16$ the Pi spends
$14\times$ as much time per puzzle as the attacker, so $d=16$ is too low for the
Pi. Each extra bit of $d$ doubles the attacker's cost, so an operator can set
$d=18$ on x86 and $d=22$ on the Pi, which costs an honest x86 client 35 or 550\,ms
(Appendix~\ref{app:puzzle-cost}).

The replay filter bounds the work per cookie as expected. The filter answers each
request of a cookie at most four times per worker (\S\ref{sec:replay}), so
floods that keep resending the requests of one cookie get few answers: such
$u$-chunk and $v$-residue floods draw 42 to 123 replies per second, and steady success is 100\,\%. A
harvesting flood, which obtains fresh cookies back-to-back, is stronger but
still weaker than a handshake flood: steady success is 69 to 100\,\%, except
from a botnet at HQC-5, where it drops to 33 and 51\,\% for $u$-chunk and
$v$-residue floods.
The attacker can also write to the filter: by resending its own requests, it
fills duplicate counters that honest requests share, and an honest
retransmission that lands on a full counter is dropped, so that client must
fetch a new cookie. Filling 39\,\% of the counters takes the attacker's full
request rate, at which it already saturates the server's CPU
(Appendix~\ref{app:replay}).
The stateful server fares far worse under the same kind of
flood. When the attacker completes key uploads, each one forces an
encapsulation: steady success is 40\,\% from one source and 4\,\% from a
botnet at HQC-5. When it only starts uploads and leaves them unfinished,
steady success is 44\,\% from one source.

Memory exhaustion is worse than CPU exhaustion for three reasons. CPU time spent
on a flood is gone once a request is answered, while a table entry stays for its
cookie's lifetime, so the attacker's work adds up. A server that runs out of
memory can crash or have other processes on the device killed, while an
overloaded CPU only slows the server, which recovers when the flood stops. And a
small device has far less memory to spare than CPU time.
Memory exhaustion targets the stateful table. Every table entry needs its own valid
cookie, so the attacker's valid cookies bound the live entries
(Fig.~\ref{fig:state}). Four Pi cores answer about 11{,}000 handshake requests
per second at every level, so with the baseline's 30\,s cookies an attacker
holds about 332{,}000 cookies, and an exhaustion-proof table needs 0.8, 1.6
and 2.5\,GB at HQC-1, 3 and 5. A smaller table is cheap to exhaust: a 25\,MB
table at HQC-1 takes only about 3\,\% of that handshake throughput and a few
thousand key upload fragments per second.
Our table-filling floods, in which each flood socket holds one table entry,
confirm this (Table~\ref{tab:tablefill} in Appendix~\ref{app:tables}): with a cap of 16, 64 or 256 entries,
steady success is 0\,\%. With 1024 entries, a flood that keeps restarting
uploads with fresh cookies leaves 0, 48 and 46\,\%; one that keeps a single
unfinished upload per socket leaves 50 to 86\,\%, only because our flood
client obtained cookies on 23\,\% of its sockets.

The stateful server can try to reduce this memory in three ways, but none makes it
independent of the attacker, so memory exhaustion remains an open attack
surface. Limiting entries to four per /64 is weak: IPv6 /48s are
routinely allocated~\cite{rfc6177}, and just two /48 subnets can hold
332{,}000 cookies.
Shorter cookies reduce the memory to 266--818\,MB at \tool's 10\,s and 80--245\,MB at
3\,s, but with 3\,s cookies up to 37\,\% of honest sessions failed on a lossy
satellite path.
A puzzle on its handshake request helps most: one attacker core
at $d=16$ then holds only 2.8 to 8.5\,MB, but an attacker can simply get more compute capacity.
\tool needs none of these mitigations; its memory is fixed at 1.2--1.6\,MiB.

\subsection{Results on Raspberry Pi 3B+}
\label{sec:eval-arm}

We answer \textbf{(Q4)} affirmatively: on the Pi, \tool keeps its lead over
McTiny, costs about as much over the stateful server as on x86, and floods
affect both designs alike (Table~\ref{tab:arm} in Appendix~\ref{app:pi}).

The low-power platform is a Raspberry Pi 3 Model B+ with four Cortex-A53 cores
and no cryptographic extensions. We fix its clock at 1.2\,GHz, because at 1.4\,GHz the
firmware automatically throttles the clock; we read the clock with \texttt{vcgencmd} during the runs.
The server runs on one pinned core. The load generator and the legitimate probes run on
the x86 host across an Ethernet LAN, whose link to the Pi carries at most 327\,Mbit/s of UDP in 1200-byte datagrams
(measured with \texttt{iperf3}).
Appendix~\ref{app:pi} gives the software and the measurement method.

The lead over McTiny holds. One \tool session costs the server 4.15, 13.87 and
31.40\,ms at HQC-1, 3 and 5; McTiny costs 283.7\,ms, $9.0\times$ more than
HQC-5. One core serves 31.8 HQC-5 sessions per second against 3.5. A client
needs 35\,ms of CPU per session at HQC-5 against a median of 7.6\,s for McTiny,
and a LAN session takes 34\,ms against 350\,ms. The server binary is 712\,KB
(McTiny: 165\,KB).

The price of statelessness is also similar to x86. The stateful server needs
2.15, 4.96 and 7.05\,ms per session, so \tool costs $1.9\times$, $2.8\times$ and $4.5\times$ as
much.
Floods on the Pi show all three attack surfaces. Without the puzzle, each flood
packet costs either server one ML-KEM decapsulation, so both cores saturate and
only 10\,\% (\tool) and 9\,\% (stateful) of probes finish within 1\,s. With the
puzzle on, \tool's core is 78 to 79\,\% busy answering about 33{,}600
challenges per second, and its memory stays fixed. Probes still fail, because
the Pi's onboard Ethernet sits on its USB~2.0 bus and receives at most
327\,Mbit/s, so the flood drops honest packets before the server sees them
(Appendix~\ref{app:pi}). On x86, where the link keeps up, the puzzle restores
full success.
Capacity grows with cores, because each worker shares
only the cookie key: in a handshake flood four workers answer $4.0\times$ as
many requests as one (11{,}186 against 2{,}786 per second at HQC-1), and the
stateful server scales by $3.9\times$. With the replay filter off, four workers
answer $3.4\times$ to $3.9\times$ as many $u$-chunk and $v$-residue requests as one.

\section{Discussion}

Stateless DoS defences for key exchange go back to Photuris
cookies~\cite{rfc2522}, JFK~\cite{jfk}, SYN cookies~\cite{rfc4987} and client
puzzles~\cite{juels_puzzles,aura_puzzles,rfc7401}; \tool preserves their core
rule that an unauthenticated peer creates no state. Fragmented post-quantum
messages suffer known failure modes~\cite{herzberg_frag,sikeridis_pqtls}; our
1232-byte limit avoids fragmentation. Our masks resemble share
re-randomization in MPC~\cite{beaver_randomization} and outsourced
computation~\cite{hohenberger_outsource}.

Our residue streaming extends to BIKE~\cite{bike_spec}, whose ciphertext
$e_0+e_1h$ is also a cyclic polynomial product with the client's key.
While an ML-KEM-768 exchange fits into a single packet each way
($14\times$ fewer bytes than \tool at category~3), streaming HQC provides
forward secrecy independent of lattice assumptions.  Alternatively, rotating
server-side ephemeral keys would avoid streaming client keys, but requires the
client's 4.4--14.4\,KB ciphertext to arrive in pieces, reintroducing fragment
buffering. QUIC~\cite{rfc9000} bounds amplification with address tokens and
padding floors, but tokens cannot stop attackers with diverse address pools. Faster HQC
multipliers~\cite{additive_fft_hqc,dong2025opthqcoptimizehqchighperformance}
could further accelerate residue evaluation.

Isogeny-based key exchange has datagram-sized public keys and needs no
streaming, but its assumptions are far younger and less studied than those of
codes and lattices: SIDH, studied for a decade, fell to a classical attack in
2022~\cite{castryck2023sidh}. CSIDH~\cite{csidh} is slow, and its quantum
security is disputed~\cite{peikert2020csieves,bonnetain2020quantum}. The recent
MIKE~\cite{mike2026} is faster but still costs the server milliseconds per
handshake, an order of magnitude more than \tool, widening the CPU-exhaustion
attack surface; its parameters already grew after a 2026
attack~\cite{wesolowski2026}, and NIST has selected neither scheme.

\tool is a key-exchange component: it ends with a session key that both sides
share, and leaves the data channel to the protocol that embeds it. Embedding it
in a full protocol is future work; as in WireGuard~\cite{wireguard}, that
protocol would create transport state only after the handshake has
authenticated the peer.
Forward secrecy in \tool rests on one family, HQC, just as it rests on lattices
alone in prior designs~\cite{pq-wireguard,rosenpass}. A simple way to rest it on
both is an extra ephemeral ML-KEM exchange at the cost of one more round trip;
more efficient methods may exist; we leave them as future work.

\section{Conclusion}
\label{sec:conclusions}

With Classic McEliece now advised against, stateless post-quantum key
exchanges that relied on it are left with lattices alone. \tool is the first
way to keep a code-based layer in such a protocol without giving up
statelessness. \tool enables a server that keeps no per-client state to accept a code-based
ephemeral key in packet-sized pieces, with a formal proof that masked replies
reveal no more than an ordinary HQC ciphertext. Compared to McTiny, it reduces
bandwidth by $27\times$ and server CPU by $9.0\times$ at category~5. Against a
stateful buffering baseline, it removes memory exhaustion, leaves CPU exhaustion to client puzzles and link
saturation to upstream filtering, and enables stateless load balancing at the
cost of moderate CPU overhead, demonstrating that
stateless post-quantum key exchange with code-based forward secrecy, built only
from NIST-selected KEMs, is practical for tiny, resource-constrained servers.

\ifarxiv\section*{Acknowledgements}
\ifarxiv
  We thank the reviewers for their feedback. This project was funded by the
  European Union under ERC Starting Grant no.\ 101115046 (``SecuStack''). Views
  and opinions expressed are however those of the authors only and do not
  necessarily reflect those of the European Union or the European Research
  Council. Neither the European Union nor the granting authority can be held
  responsible for them.
\else
  Redacted for anonymous submission.
\fi
\fi

\bibliographystyle{IEEEtran}
\bibliography{references}

\appendices
\section{Wire Format and Key Schedule}
\label{app:wire}

Every packet starts with a two-byte header. In byte~0, bit~7 is the direction (1 for
server to client), bit~5 is the cookie key index $b$, bits~0--2 are the kind
(0 handshake, 1 $u$-chunk, 2 $v$-residue, 3 challenge, 4 solved handshake), and all other bits are zero. Byte~1 is the
index: the chunk number $j$ or the residue number $i-1$ (0 for the handshake). Layouts, in bytes:
\begin{center}\scriptsize
\begin{tabular}{@{}lp{0.68\columnwidth}@{}}
Handshake request & hdr 2 $|$ $\mathrm{hash}(pk)$ 32 $|$ $\mathit{seed}_h$ 32 $|$ $c_{\mathrm{auth}}$ 1{,}088 \\
Handshake reply & hdr $|$ nonce 24 $|$ AE(cookie $|$ salt 16 $|$ $\gamma$ 32) $|$ tag 16 \\
$u$-chunk request & hdr $|$ cookie $|$ tag 16 $|$ zeros \\
$u$-chunk reply & hdr $|$ nonce 24 $|$ AE($u$ bytes, at most 1190) $|$ tag 16 \\
$v$-residue request & hdr $|$ cookie $|$ AE($s_i$) $|$ tag 16 \\
$v$-residue reply & hdr $|$ nonce 24 $|$ AE($v_i$, $\lceil\deg f_i/8\rceil$) $|$ tag 16 \\
Cookie & nonce 24 $|$ AE($\theta$ 32 $|$ $m$ $|$ $\mathit{seed}_h$ 32 $|$ $K_{\mathrm{auth}}$ 32) $|$ tag 16 \\
\end{tabular}
\end{center}
The message $m$ has 16, 24 or 32 bytes at HQC-1, 3 and 5, so the cookie has 152,
160 or 168 bytes. The sealed
residue $s_i$ has 884, 996 or 1030 bytes, zero-padded to the longest residue of the level. The
associated data of a request is its header and cookie; that of a reply is its header,
plus the BLAKE3 digest of the request body for the handshake reply. The cookie
has empty associated data. The server accepts exactly one length per request kind and
rejects non-zero padding, non-zero unused bits of a residue, reserved bits and indices out of range.

\emph{Key schedule.} The keys $k_{c\to s}$ and $k_{s\to c}$ are BLAKE3 \texttt{derive\_key} of
$K_{\mathrm{auth}}$ under two distinct context strings. The joint secret is
\texttt{derive\_key}(context, $|K|\,\|\,K\,\|\,K_{\mathrm{auth}}$), and
$K_{\mathrm{session}}$ and $K_{\mathrm{trans}}$ are two further derivations of it. The
tag $\gamma$ is a keyed BLAKE3 hash under $K_{\mathrm{trans}}$ of the digest of the
request, the salt and the key index. The quotient mask $\mu$ is a ChaCha12 stream under a key equal to the BLAKE3
hash of $\theta$, the context string \texttt{crt quotient mask} and a label. The
truncate mask $\tau$ is read from the BLAKE3 extendable output of $\theta$, the
context string \texttt{crt truncate mask} and a label.

\section{Sizing the Replay Filter}
\label{app:replay}

Let $R$ be the maximum number of requests a worker answers per second and $R_c=10$\,s
the period of the cookie-key rotation. The filter starts a new generation at each
rotation and keeps the two newest. An entry thus stays until the second rotation
after it was written. The cookie it belongs to can no longer be opened at that
point, because a cookie opens only until the second rotation after it was issued.
One filter generation must hold $n=R\cdot R_c$ fingerprints. For a target
false-positive rate $p$, a Bloom filter needs $m=\lceil -n\ln p/(\ln 2)^2\rceil$
bits and $k=\mathrm{round}((m/n)\ln 2)$ hash functions; at $p=0.001$ this is
14.4 bits per fingerprint and $k=10$. The duplicate counters add $\lceil n/2\rceil$
slots of 2 bits. Both generations together need twice this. We set $R$ to the
inverse of the Criterion cost of the cheapest accepted request, the $v$-residue
request, rounded down to a multiple of 100: on x86 with the filter on
(31.93, 46.32 and 65.31\,\textmu s), on the Pi from the handler costs of
Table~\ref{tab:arm}. A query searches
both generations, so the worst-case false-positive rate is about $2p$; a false
positive is still answered unless its duplicate counter is full.
A separate hash of the fingerprint picks the counter slot, so about two
fingerprints share a slot. An attacker that resends each of its own requests $D$
times fills the slots of its fingerprints. If it spends a fraction $f$ of $R$ on
this, it fills about $1-e^{-2f/(D+1)}$ of the slots, 39\,\% at $f=1$ and $D=3$. A
first request is always fresh, so only an honest resend whose request reached the
server can be dropped. It is then dropped on every retry, and the client must
fetch a new cookie. At $f=1$ the attacker already saturates the server's CPU
(\S\ref{sec:eval-dos}).

\begin{table}[H]
\caption{Replay-filter size per worker ($p=0.001$, $R_c=10$\,s).}
\label{tab:replay}
\centering\footnotesize\setlength{\tabcolsep}{3pt}
\begin{tabular}{@{}lrrrr@{}}\toprule
Host, level & $R$ (1/s) & $n$ & $m$ (bits) & Memory \\ \midrule
x86, HQC-1 & 31{,}300 & 313{,}000 & 4{,}500{,}185 & 1175.1\,KiB \\
x86, HQC-3 & 21{,}500 & 215{,}000 & 3{,}091{,}182 & 807.2\,KiB \\
x86, HQC-5 & 15{,}300 & 153{,}000 & 2{,}199{,}771 & 574.4\,KiB \\
Pi, HQC-1 & 2{,}400 & 24{,}000 & 345{,}063 & 90.1\,KiB \\
Pi, HQC-3 & 1{,}400 & 14{,}000 & 201{,}287 & 52.6\,KiB \\
Pi, HQC-5 & 800 & 8{,}000 & 115{,}021 & 30.0\,KiB \\ \bottomrule
\end{tabular}
\end{table}

\section{CRT Moduli}
\label{app:moduli}

\begin{table}[H]
\caption{CRT moduli table; $j\times d$ means $j$ moduli of degree $d$.}
\label{tab:moduli}
\centering\footnotesize\setlength{\tabcolsep}{4pt}
\begin{tabular}{@{}lrrr@{}}\toprule
 & HQC-1 & HQC-3 & HQC-5 \\ \midrule
$k$ & 5 & 9 & 14 \\
$j\times d$ & $2{\times}7068$, $3{\times}7067$ & $7{\times}7967$, $2{\times}7966$ & $11{\times}8234$, $3{\times}8233$ \\
$2n-1$ & 35{,}337 & 71{,}701 & 115{,}273 \\ \bottomrule
\end{tabular}
\end{table}

\section{x86 Testbed}
\label{app:x86}

All x86 results come from one AMD Ryzen 7 5800H with AVX2, boost off and the
\texttt{performance} governor. The pinned cores ran at 3169.1\,MHz (largest
deviation 0.006\,\%). The server runs on cpu0, whose hyperthreading sibling core
is offline; the load generator and the session client run on cpu2 and the
legitimate probes on cpu4. \tool and the baseline run in a Docker image (Rust
nightly 2026-06-15); McTiny on the host (Ubuntu 26.04, Linux 7.0, Docker 29.8).
Cycles are computed as elapsed time multiplied by 3169.1\,MHz. Network
emulation uses \texttt{tc netem} with half of the delay on each side.

\section{Raspberry Pi Details}
\label{app:pi}

\begin{table}[H]
\caption{Raspberry Pi 3B+, one worker. Success: probes done within 1\,s / 5\,s. $^\dagger$Replay filter off.}
\label{tab:arm}
\centering\scriptsize\setlength{\tabcolsep}{3pt}
\begin{tabular}{@{}lrrrr@{}}\toprule
 & HQC-1 & HQC-3 & HQC-5 & McTiny \\ \midrule
Session CPU, \tool (ms) & 4.15 & 13.87 & 31.40 & 283.7 \\
Session CPU, stateful (ms) & 2.15 & 4.96 & 7.05 &  \\
McTiny / \tool & --- & --- & 9.0$\times$ &  \\
Handshake request (ms) & 0.301 & 0.303 & 0.302 & 2.65 \\
$u$-chunk request (ms) & 0.538 & 1.545 & 1.904 &  \\
$v$-residue request (ms) & 0.416 & 0.695 & 1.129 &  \\
Capacity (derived), sessions/s & 241.1 & 72.1 & 31.8 & 3.52 \\
Capacity (derived), queries/s & 1{,}929 & 1{,}010 & 701 & 3{,}418 \\
Flood: handshakes answered/s & 2{,}786 & 2{,}755 & 2{,}761 & --- \\
Flood: $u$-chunks answered/s$^\dagger$ & 1{,}480 & 544 & 434 & --- \\
Flood: $v$-residues answered/s$^\dagger$ & 1{,}833 & 1{,}175 & 734 & --- \\
Flood / derived, $v$-residue (\%)$^\dagger$ & 76 & 82 & 83 & --- \\
Legit. success, handshake flood (\%) & 10/40 & --- & 10/10 & --- \\
Same, puzzle $d{=}16$ (\%) & 10/40 & --- & 10/20 & --- \\
Same, stateful server (\%) & 9/36 & --- & 9/9 & --- \\
Client CPU per session (ms) & 19 & 27 & 35 & 7{,}578 \\
LAN session time (ms) & 7 & 16 & 34 & 350 \\
Minimum stack (KiB) & 96 & 80 & 164 & 64 \\
\bottomrule\end{tabular}\end{table}

\emph{Software.} The HQC polynomial multiplier is the one of Chen, Chiu, Peng and
Yang~\cite{additive_fft_hqc}: Kronecker substitution with a Chinese-remainder step (KS+CRT) that uses the
8-bit polynomial multiply \texttt{PMULL.P8}, for HQC-1 and HQC-5, and their Toom--Karatsuba port
for HQC-3. For HQC-3 we do not use KS+CRT because its multiplier reads stack
memory after a stack-pointer reset at this size (found with Valgrind). The rest of HQC
is the reference implementation. Our own multiplication modulo the trinomials uses the NEON
instruction \texttt{vmull\_p8}~\cite{camara2013}. ML-KEM-768 is
libcrux-ml-kem~\cite{libcrux}. McTiny uses the \texttt{vec} implementation of
libmceliece 20230612~\cite{libmceliece} on the Pi.
Valgrind on the Pi finds no secret-dependent branch or address in the server paths, ML-KEM-768
decapsulation or multipliers, only branches on public values (tag checks and the check that
a client's residue is reduced) and HQC's rejection sampler in client key generation.

\emph{Method.} We measure the handler cost per request with Criterion, the server CPU per
session with \texttt{schedstat}, the client costs, the memory and the session time over the LAN.
The live flood runs from the x86 host, with the replay filter on. Probes start every 1\,s during 10\,s floods (HQC-1) or every 4\,s during 40\,s floods (HQC-5), and every 200\,ms in the 10\,s harvesting floods, with and without the puzzle. We count the server's replies with the kernel's UDP send counter. McTiny's denial-of-service capacity on the Pi is projected from its measured per-session CPU cost, as no packet replay utility exists for its fragmented wire format. We apply the same projection method to \tool
and validate it against live LAN floods of one \tool worker: it answered 76 to 84\,\% of the rate projected from handler costs, with the remainder consumed by kernel network ingress. The $u$-chunk and $v$-residue measurements disable the replay filter to assess worst-case computation per request.
Table~\ref{tab:arm} gives the results.

\emph{Live floods.} With the puzzle on, the Pi's network ingress limits the
server: at $d=16$ it answers 33{,}600 to 33{,}700 challenges per second with its core 78 to
79\,\% busy, while the flood sends about 160{,}000 packets per second, and 10\,\% of probes
finish within 1\,s and 20 to 40\,\% within 5\,s. The Pi's onboard Ethernet sits on its
USB~2.0 bus, a limit that would affect any UDP service on this board. Floods of $u$-chunk and
$v$-residue requests that reuse one cookie hit a similar limit: the filter lets through only
6 to 11 answers per second, but the large flood packets fill the link, and 25 to 50\,\% of
probes finish within 5\,s.

\section{Flood and Latency Tables}
\label{app:tables}
\label{app:botnet}

Tables~\ref{tab:flood} and~\ref{tab:flood-botnet} give the flood results of \S\ref{sec:eval-dos} at every level, for one source and for 1024 distinct source addresses on our testbed. Table~\ref{tab:tablefill} reports the state table filling experiments, and Table~\ref{tab:wan} the session times on emulated paths.

\emph{Flood method.} We flood the server for 10 seconds after client set-up, launching one legitimate
probe every 200\,ms from an independent core (10 repetitions per configuration, 5 for harvesting floods).
The \emph{single} shape sends from 128 sockets of one address; the \emph{botnet}
shape uses 4096 sockets across 1024 loopback addresses to emulate distributed traffic. Handshake floods carry either valid or invalid (\texttt{auth}) ciphertexts $c_{\mathrm{auth}}$, with the replay filter enabled.
A $u$-chunk or $v$-residue flood first completes an initial handshake to obtain a valid cookie and requests for each index. In a \emph{cycling} flood, the client loops through these requests, renewing the handshake every 5\,s. In a \emph{harvesting} flood, the client executes handshakes back-to-back across all sockets, exhausting the replay filter limit for every cookie to maximize server work.
Floods against the stateful baseline hold a cookie per socket: \texttt{partial} sends a first fragment and leaves the upload unfinished, \texttt{full} uploads the whole key, so that each completed upload forces an encapsulation and frees its entry for the next upload, and \texttt{churn2} keeps restarting each socket with a new cookie and an unfinished upload.
We evaluate two metrics: \emph{steady success} (probes started during steady-state flooding that finish within 1\,s) and \emph{done in flood} (the fraction of started probes that complete before the flood ends), reporting medians across repetitions.

\begin{table}[t]
\caption{Single-source floods, one server core: steady success / done in flood (\%) for \tool (ours) and the stateful baseline (base). $d{=}16$: puzzle on.}
\label{tab:flood}
\centering\scriptsize\setlength{\tabcolsep}{2pt}
\begin{tabular}{@{}lrrrrrr@{}}\toprule
 & \multicolumn{2}{c}{HQC-1} & \multicolumn{2}{c}{HQC-3} & \multicolumn{2}{c}{HQC-5} \\
Flood (ours/base) & ours & base & ours & base & ours & base \\ \midrule
handshake & 52/78 & 23/41 & 42/74 & 15/25 & 32/70 & 12/20 \\
handshake, $d{=}16$ & 100/100 & --- & 100/100 & --- & 100/100 & --- \\
$u$-chunk, cycling & 100/100 & --- & 100/100 & --- & 100/100 & --- \\
$v$-residue, cycling & 100/100 & --- & 100/100 & --- & 100/100 & --- \\
$u$-chunk, harvesting & 100/100 & --- & 99/100 & --- & 97/100 & --- \\
$v$-residue, harvesting & 100/100 & --- & 100/100 & --- & 99/100 & --- \\ \midrule
---/auth & --- & 16/42 & --- & 18/34 & --- & 8/18 \\
---/partial & --- & 82/94 & --- & 76/87 & --- & 44/43 \\
---/full & --- & 64/84 & --- & 56/68 & --- & 40/46 \\
\bottomrule\end{tabular}\end{table}

\begin{table}[H]
\caption{Botnet floods, one server core: steady success / done in flood (\%), as in Table~\ref{tab:flood}. \tool rows: 5 runs; \texttt{partial}: same runs as Table~\ref{tab:tablefill}.}
\label{tab:flood-botnet}
\centering\scriptsize\setlength{\tabcolsep}{2pt}
\begin{tabular}{@{}lrrrrrr@{}}\toprule
 & \multicolumn{2}{c}{HQC-1} & \multicolumn{2}{c}{HQC-3} & \multicolumn{2}{c}{HQC-5} \\
Flood (ours/base) & ours & base & ours & base & ours & base \\ \midrule
$u$-chunk, harvesting & 100/100 & --- & 69/100 & --- & 33/98 & --- \\
$v$-residue, harvesting & 97/100 & --- & 78/100 & --- & 51/98 & --- \\ \midrule
---/handshake & --- & 18/39 & --- & 16/34 & --- & 10/22 \\
---/auth & --- & 12/38 & --- & 16/34 & --- & 10/20 \\
---/partial & --- & 86/93 & --- & 74/81 & --- & 50/48 \\
---/full & --- & 4/18 & --- & 6/16 & --- & 4/12 \\
\bottomrule\end{tabular}\end{table}

\begin{table}[H]
\caption{Stateful server under table-filling floods, by table cap (steady success within 1\,s, \%).}
\label{tab:tablefill}
\centering\scriptsize\setlength{\tabcolsep}{2.5pt}
\begin{tabular}{@{}rrrrrrrr@{}}\toprule
 & \multicolumn{3}{c}{\texttt{partial}} & \multicolumn{3}{c}{\texttt{churn2}} & \texttt{churn2} \\
Cap & HQC-1 & HQC-3 & HQC-5 & HQC-1 & HQC-3 & HQC-5 & HQC-5 MB \\ \midrule
16 & 0 & 0 & 0 & 0 & 0 & 0 & 0.12 \\
64 & 0 & 0 & 0 & 0 & 0 & 0 & 0.47 \\
256 & 0 & 0 & 0 & 0 & 0 & 0 & 1.89 \\
1024 & 86 & 74 & 50 & 0 & 48 & 46 & 7.57 \\
\bottomrule\end{tabular}\end{table}

\begin{table}[t]
\caption{Median session time at HQC-5 on emulated paths (s). Ratio: McTiny / \tool, from unrounded medians; under loss it mostly reflects McTiny's 120\,s timer.}
\label{tab:wan}
\centering\scriptsize\setlength{\tabcolsep}{3pt}
\begin{tabular}{@{}lrrrrr@{}}\toprule
Cell (RTT, loss/direction) & \tool & p95 & Stateful & McTiny & Ratio \\ \midrule
2\,ms, 0\,\% & 0.007 & 0.008 & 0.006 & 0.069 & 9.7 \\
20\,ms, 0.05\,\% & 0.046 & 0.049 & 0.046 & 0.414 & 8.9 \\
75\,ms, 0.1\,\% & 0.166 & 0.325 & 0.164 & 0.999 & 6.0 \\
140\,ms, 0.2\,\% & 0.306 & 0.868 & 0.304 & 1.82 & 5.9 \\
300\,ms, 0.5\,\% & 0.636 & 1.832 & 0.637 & 3.78 & 5.9 \\
20\,ms, 0\,\% & 0.045 & 0.048 & 0.046 & 0.279 & 6.2 \\
20\,ms, 1\,\% & 0.049 & 0.281 & 0.047 & 1.12 & 22.6 \\
20\,ms, 2\,\% & 0.165 & 0.430 & 0.048 & 3.69 & 22.4 \\
20\,ms, 2.5\,\% & 0.107 & 0.172 & 0.047 & 7.14--7.33 & 67--69 \\
20\,ms, 3\,\% & 0.164 & 0.173 & 0.148 & 14.1--51.6 & 86--314 \\
20\,ms, 4\,\% & 0.168 & 0.406 & 0.144 & 77.8--103.1 & 464--615 \\
20\,ms, 5\,\% & 0.169 & 0.336 & 0.276 & 120.7--121.3 & 713--717 \\
20\,ms, Gilbert--Elliott & 0.170 & 0.463 & 0.049 & 120.7--120.8 & 712 \\
\bottomrule\end{tabular}\end{table}

\section{Puzzle Cost Model}
\label{app:puzzle-cost}

With the puzzle on, a cookie costs the attacker $2^d$ hashes on average, 8.7\,ms at
$d=16$ on one x86 core. A solved puzzle buys at most $D+1$ decapsulations, all with one
cookie (\S\ref{sec:puzzle}), and the replay filter answers each request of that cookie at
most $D+1$ times per worker. One puzzle therefore buys at most $W=(D+1)c_s$ per worker,
where $c_s$ is the server CPU per session (Tables~\ref{tab:cpu} and~\ref{tab:arm}): 8.3\,ms
on x86 and 126\,ms on the Pi at HQC-5. The worst-case ratio of server to attacker time is
then 0.95 on x86 and 14 on the Pi, times $N$ for $N$ workers, and each bit of $d$ halves it.
The buffering baseline as built has no such bound: a completed upload frees its entry, so
one cookie buys an encapsulation per repeated upload. Given the same puzzle and $D+1$
limits, its bound would be $D+1$ sessions, 2.0\,ms at HQC-5, so statelessness costs
$\log_2(8.3/2.0)\approx2.0$ bits of $d$. On the Pi, a harvester that solves every puzzle at
$d=16$ got 7 to 11 answers per cookie, and half an attacker core kept the server core 73 to
82\,\% busy at HQC-3 and 5, a ratio of 1.2 to 1.8, below the worst case.

\section{Ethics Considerations}
\label{app:ethics}

All experimental evaluations and flooding attacks were conducted strictly on isolated, private laboratory networks on hardware that we own and operate. No production networks, third-party infrastructure, or external systems were targeted. Our experiments evaluated only our own prototype implementation and the published research artifact of McTiny~\cite{mctiny}. Implementation defects identified during testing (such as the stack issue in an upstream research multiplier noted in Appendix~\ref{app:pi}) were handled responsibly and circumvented using alternative routines.

\end{document}